\documentclass[%
 reprint,
 superscriptaddress,
 nofootinbib,
 amsmath,amssymb,
 aps,
]{revtex4-2}

\usepackage{graphicx}
\usepackage{dcolumn}
\usepackage{bm}
\begin{document}

\preprint{APS/123-QED}

\title{Gravitational wave signals in the deci-Hz range from neutrinos during the proto-neutron star cooling phase}
\author{Lei Fu}\email{fulei@akane.waseda.jp}
\affiliation{%
 Graduate School of Mathematics, Nagoya University, Furocho, Chikusaku, Nagoya, 464-8602, Japan 
}%
\author{Shoichi Yamada}
\affiliation{%
 Department of Science and Engineering, Waseda University, 3-4-1 Okubo, Shinjuku, Tokyo 169-8555, Japan 
}%

\date{\today}
\begin{abstract}
We investigate the gravitational waves (GWs) at low frequencies produced by neutrinos that are emitted anisotropically from the proto-neutron star (PNS) during its cooling phase that lasts for about a minute. We are particularly interested in the deci-Hz range, to which some satellite-borne detectors are expected to have good sensitivities. We first give a formulation based on the spherical-harmonic expansion of the neutrino luminosity to obtain the gravitational waveform as well as the characteristic strain. In the absence of multi-dimensional simulations of PNS cooling, from which we can extract reliable data on the neutrino luminosities as a function of solid angle, we construct them by hand. In the first model, the time evolution is approximated by piece-wise exponential functions (PEFs); in the second model we employ the time profile obtained in a 1D cooling simulation for all harmonic components for simplicity; 
In both cases, we consider not only axisymmetric components but also non-axisymmetric ones; as the third model, we consider axisymmetric neutrino emissions, the axis of which is misaligned with the rotation axis and, as a result, rotates with the PNS. We find from the first model that the decay times in PEF at late phases can be inferred from the positions of bumps and dips in the characteristic strain of the GW in the case of a slow cooling, whereas they may be obtained by identifying the positions of slope change in the case of rapid cooling, which may be induced by convections in PNS. We confirm the former result also in the second model.
The results of the third model show that the gravitational waves emitted by the neutrinos contain circularly polarized components in contrast to the first two models, in which only linear polarizations occur, and give oscillatory features in the waveform with the frequencies of integral multiples of the rotation frequency, which are also clearly reflected in the characteristic strain.
Finally we compare the GW signals thus obtained with the sensitivity curves of some planned GW detectors that are expected to be sensitive to GWs in the deci-Herz range. If these admittedly crude models are any guide, these detectors, DECIGO in particular, will have a fair chance to detect the GW signals as we consider them in this paper if they are emitted from within our Galaxy.
\begin{description}
\item[Usage]
Publication
\item[Keywords]
Gravitational waves - neutrinos - proto neutron star
\end{description}
\end{abstract}

\maketitle


\section{\label{sec:level1}Introduction}
Core-collapse supernovae (CCSNe)\cite{Burrows_2021} are explosion phenomena that occur at the final stage of massive stars. In this process, neutron stars\cite{Janka2001} are produced as results normally (see, e.g., \cite{O_Connor_2011} for possible black hole formations). 
After the launch of an explosion, the central region filled with neutrons is very hot, $T \gtrsim 10^{11}$ K, and still proton- and lepton-rich, $Y_{p}\sim Y_{e} \gtrsim 0.2$\cite{2017}. It is hence referred to as the proto-neutron star (PNS)\cite{Prakash20011}.
It cools and deleptonizes by emitting neutrinos copiously for about a minute\cite{Roberts_2012}\cite{Roberts_2017}\cite{Nakazato_2020}. This stage is called the cooling phase of PNS. 
This is probably the most important stage for neutrino emissions from an observational  point of view. In fact, what Kamiokande and IMB detectors observed from SN1987A in the Large Magellanic Cloud is the neutrinos produced in this phase.
In this paper, we will pay attention to the gravitational waves (GWs) that those neutrinos emit during the same phase. We are particularly concerned with the sub-Hz range. This may be understandable if one is reminded that the duration of the cooling phase is about a minute. It has been known that emissions/absorptions/scatterings of particles are accompanied by the emission of GWs in general\cite{Weinberg:100595}\cite{1977ApJ...216..610T}. It was Epstein\cite{osti_6879389} that first considered as a promising GW source asymmetric emissions of a copious amount of neutrinos from astronomical events such as CCSNe. It was Burrows and Hayes\cite{1996Pulsar}, though, that first addressed this issue based on realistic simulations of CCSNe. \\
CCSNe had been considered to be a promising GW source and studied extensively\cite{Ott_2009}\cite{Kotake_review} much before the first direct detection of GWs from a coalescence of a black hole binary by LIGO in 2016\cite{2016GW}. 
In these investigations, the GW emissions from neutrinos were indeed calculated quantitatively in addition to those from non-spherical matter motions, which will occur commonly via hydrodynamical instabilities, rotation, and magnetic stresses\cite{M_ller_2012}\cite{Mueller1997GravitationalRF}\cite{Kotake_2009}\cite{Vartanyan_2020}\cite{Favata_2010}. They found that the GWs from the neutrinos emitted anisotropically have the following properties: (1) their typical frequencies are lower than those of the GWs produced by matter, since the neutrino luminosities change more gradually; (2) the GW signal from neutrino has a so-called memory, i.e., the metric perturbation does not go back to zero after the passage of GW; in other words, the zero frequency limit of the GW signal is nonvanishing; (3) the GW strain may be dominated by the contribution from neutrinos at low frequencies although the energy carried by those GWs is minor.
See \cite{Ott_2009}\cite{Kotake_review} for more details.\\
These studies focused on the GW emissions during the post-bounce phase of CCSNe, i.e., up to $\sim$1 sec after the core bounce. More recently, Mukhopadhyay et al.\cite{Mukhopadhyay_2021} extended the exploration to the PNS cooling phase. They approximated the neutrino light curve with an exponentially decaying function with a single decay constant; the anisotropy of the neutrino emission is specified by a single parameter, which is assumed to be a multi-Gaussian function of time. They confirmed that the GW signals have  memories, which correspond to the zero frequency limit of the characteristic strain. Comparing these signals with its sensitivity, they demonstrated that DECIGO, a planned satellite-borne GW detector, will be able to detect them if they are emitted from a source at the distance of 10kpc.\\
In this paper, instead of using the single parameter to characterise the anisotropy, we employ the spherical-harmonic expansion of the neutrino luminosity, which enables us to perform angular integrations exactly; the time evolution of the coefficient for each harmonic component is modeled in three ways up to 50s. We then calculate the GW waveforms as well as their characteristic strains for these models. We also investigate the polarizations and the anisotropies of the GW emissions. 
In one of the three cases, we consider a rotating neutrino beam, which may occur in a rotating magnetized PNS with the magnetic axis misaligned with the rotation axis. We will see that the rotation frequency is reflected in the characteristic strain.\\
In order to see the detectability, we simply compare the characteristic strains obtained in our models with the planned sensitivities of near-future detectors, LISA\cite{baker2019laser}, DECIGO\cite{2001}\cite{2011}, ALIA\cite{Gong_2015} that are sensitive to the deci-hertz GW signals. Considering the simplicity of our models, we do not think it is needed to elaborate on the assessment of the detectability further, with noises taken into account properly. The results of this simple comparison indicate that all these detectors, DECIGO in particular, will have a good chance to detect the GW signals if they are emitted from a distance of 10kpc unless our expectation of the neutrino asymmetry is not widely off the mark.\\
This paper is organized as follows. Our formulation with the spherical harmonics expansion as well as the three models are described in section 2. In section 3, we present the main results and analyses thereof. Finally, we close the paper with the conclusions and some prospects in section 4.

\section{\label{sec:level1}Methods and models}
\subsection{Formulation}
\noindent The facts that the emissions, absorptions and scatterings of neutrinos (actually of any particles) are accompanied by the productions of gravitational waves (GWs) and that the phenomenon may be important in astronomical events like supernovae were first pointed out by Epstein\cite{osti_6879389} and Turner\cite{Turnerrr}. Burrows \& Hayes\cite{1996} were the first to apply the formula to a realistic supernova model and touched the GW memory.
More systematic investigations were conducted by Mueller et al.\cite{M_ller_2012}\cite{Mueller1997GravitationalRF}. The formula they developed will be extended in the following
(see also \cite{Kotake_2009}\cite{Vartanyan_2020}\cite{Mukhopadhyay_2021}).\\
The gravitational wave that is emitted from the neutrinos that are radiated anisotropically from a point source is given as 
\begin{equation}
    h_{S}(t,\alpha,\beta)=\frac{2G}{c^{4}R}\int^{t}_{0}dt' \int d\Omega \,W_{S}(\Omega,\alpha,\beta)\frac{dL_{\nu}}{d\Omega}(\Omega,t'),
\end{equation}
where $S\in(+,\times)$ specifies the GW polarizarion, and $dL_{\nu}/d\Omega$ is the neutrino energy radiated per time and per solid angle; $R$ is the distance from the source to the observer; $W_{S}(\theta, \phi, \alpha, \beta)$ is the geometric weight given by
\begin{equation}
    W_{S}(\theta,\phi,\alpha,\beta)=\frac{D_{S}(\theta,\phi,\alpha,\beta)}{N(\theta,\phi,\alpha,\beta)},
\end{equation}

\begin{align}
    D_{+}=&[1+(\cos\phi\cos\alpha+
    \sin\phi\sin\alpha)\sin
    \theta\sin\beta+\cos\theta\cos\beta]\notag\\
    &([(\cos\phi\cos\alpha+\sin\phi\sin
    \alpha)\sin\theta\cos\beta-\cos
    \theta\sin^{2}\beta]\notag\\
    &-\sin^{2}\theta(\sin^{2}\cos
    {\alpha-\cos\phi
    \sin\alpha}^{2})),
\end{align}
\begin{align}
    D_{\times}=&[1+(\cos\phi\cos\alpha+\sin\phi\sin\alpha)\sin\theta\sin\beta+\cos\theta\cos\beta]\notag\\
    &2[(\cos\phi\cos\alpha+\sin\phi\sin\alpha)\sin\theta\cos\alpha-\cos\theta\sin\beta]\notag\\
    &\sin\theta(\sin\phi\cos\alpha-\cos\phi\sin\alpha),
\end{align}
\begin{align}
    N=&[(\cos\phi\cos\alpha+\sin\phi\sin\alpha)\sin\theta\cos\beta-\cos\theta
    \sin\beta]^{2}\notag\\
    &+\sin^{2}\theta(\sin\phi\cos\alpha-\cos\phi\sin\alpha)^{2},
\end{align}
where $\theta \in[0, \pi]$ and $\phi \in [0, 2\pi]$ are the zenith and azimuth angles to specify $\Omega$ whereas $\beta \in [0,\pi]$ and $\alpha \in [-\pi, \pi]$ are the zenith and azimuth angles to give the direction to the observer (see Fig.~1). Note that Eq.~(1) is actually applicable to a source of a finite size as long as the observer is located at a sufficiently large distance from the source.
\begin{figure}
\centering 
\includegraphics[width=0.5\textwidth]{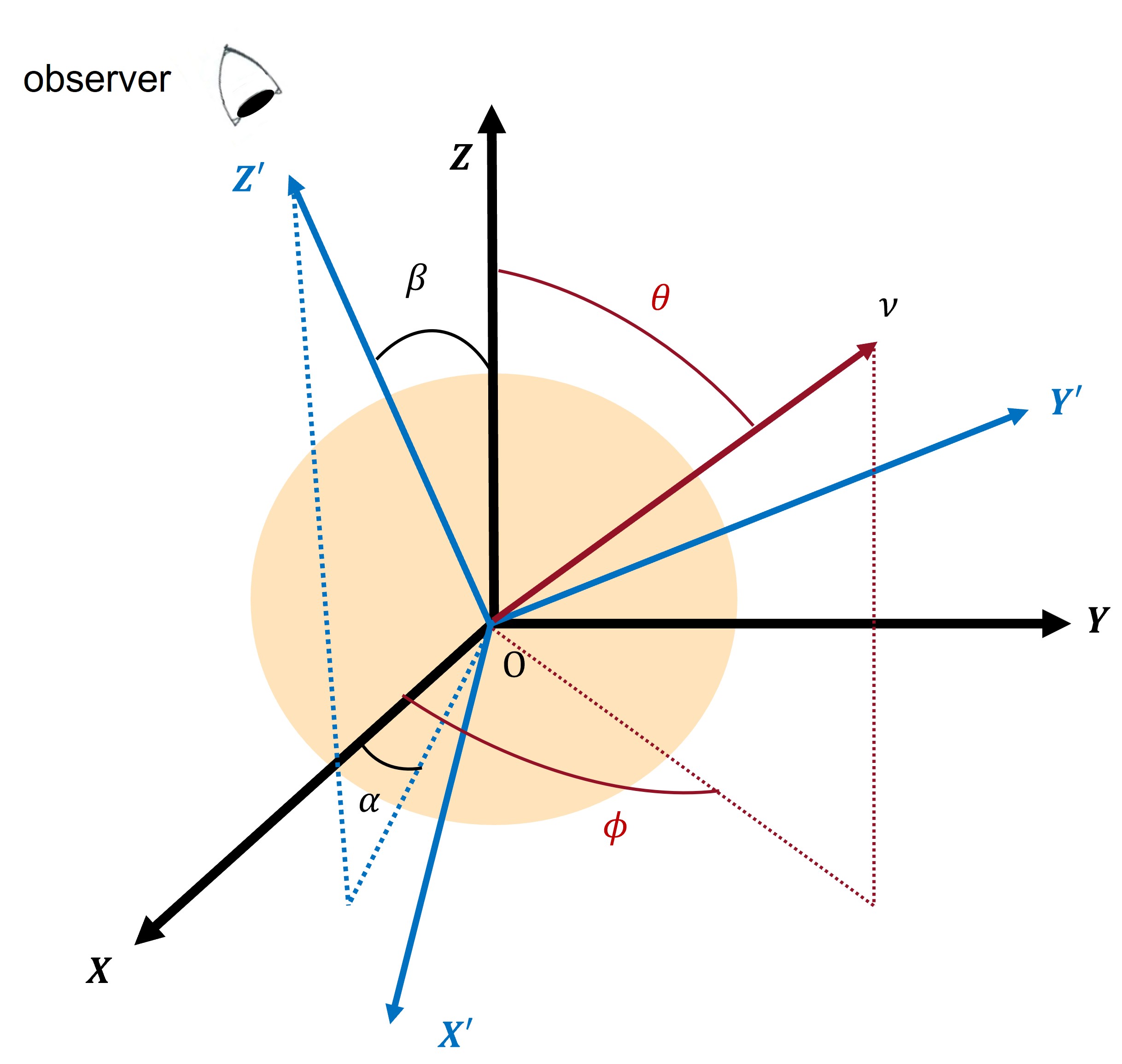}
\caption{The source coordinates ($\it X$, $\it Y$, $\it Z$) and the observer coordinates ($\it X'$, $\it Y'$, $\it Z'$). The observer is located at a large distance on the positive $\it Z'$-axis. The zenith and azimuth angles $\beta$ and $\alpha$ thus specify the direction of the observer. The orange sphere is the PNS.}
\end{figure}

\noindent The anisotropy of neutrino emissions may be expressed most conveniently with the spherical-harmonic expansion. Since the GW production is essentially governed by linear equations for astronomical events of our concern, each harmonic component can be treated separately. From an observational point of view, it is true that not individual spherical-harmonic components but the total signal is all we obtain. From a theoretical point of view, which we take in this paper, however, the total signal cannot be understood unless we
understand its individual harmonic components in detail. In addition, as will be shown shortly, the spherical-harmonic expansion enables us to perform angular integrations in advance and to understand the angular dependence of GW emissions.\\
In this paper we do not employ the ordinary complex spherical harmonic functions given as
\begin{equation}
Y^{m}_{l}(\theta,\phi)=C_{l}^{m}P_{l}^{m}(\cos\theta)e^{im\phi},
\end{equation}
where $P^{l}_{m}(\cos\theta)$ is the associated Legendre polynomial $(l = 0, 1, 2, ...$ and $m = -l, -l+1,..., l)$ and
$C_{l}^{m}$ is the normalization factor; instead we utilize the real harmonics defined  as:
\begin{equation}
\boldsymbol{Y}^{m}_{l}(\theta,\phi)= \left\{
\begin{array}{rcl}
\frac{1}{i\sqrt{2}}(Y^{m}_{l}(\theta ,\phi)-Y^{-m}_{l}(\theta ,\phi)) & & m<0\\
Y^{m}_{l}(\theta ,\phi)& & m=0\\
\frac{1}{\sqrt{2}}(Y^{m}_{l}(\theta ,\phi)+Y^{-m}_{l}(\theta ,\phi))& & m>0
\end{array} \right.
\end{equation}
or more concretely given  as
\begin{equation}
\boldsymbol{Y}^{m}_{l}(\theta,\phi)
=\left\{
\begin{array}{rcl}
\sqrt{2}K^{m}_{l}P^{m}_{l}(\cos\theta)\cos(m\phi) & & m<0\\
K^{m}_{l}P^{m}_{l}(\cos\theta)  & & m=0\\
\sqrt{2}K^{m}_{l}P^{m}_{l}(\cos\theta)\sin(m\phi)  & & m>0
\end{array} \right.
\end{equation}
with
\begin{equation}
    K^{m}_{l}=(-1)^{\frac{m+|m|}{2}}\sqrt{\frac{2l+1}{4\pi}\frac{(l-m)!}{(l+m)!}}.
\end{equation}
Then we expand  $dL(t,\theta,\phi)/d\Omega$ as 
\begin{align}
    \frac{dL(t,\theta,\phi)}{d\Omega}&=\sum^{\infty}_{l=0}\sum^{l}_{m=-l}\frac{a_{lm}(t)}{4\pi}\boldsymbol{Y}^{m}_{l}(\theta,\phi).
\end{align}
Inserting the above expression into Eq.~(1), we obtain the following formula for the gravitational waveform:
\begin{align}
    h_{S}(t,\alpha,\beta)&=\sum^{\infty}_{l=0}\sum^{l}_{m=-l}\frac{2G}{c^{4}R}\int_{0}^{t}dt'a_{lm}(t')dt'\notag\\
    &\times \frac{1}{4\pi}\int_{\Omega}d\Omega\, W_{S}(\theta,\phi,\alpha,\beta) \boldsymbol{Y}^{m}_{l}(\theta,\phi)\notag\\
    &=\sum^{\infty}_{l=1}\sum^{l}_{m=-l}h^{\rm amp}_{lm}(t)\Psi_{lm}^{S}(\alpha,\beta),
\end{align}
with
\begin{align}
    &h^{\rm amp}_{lm}(t)=\frac{2G}{c^{4}R}\int_{0}^{t}dt'a_{lm}(t')dt',\\
    &\Psi_{lm}^{S}(\alpha,\beta)=\frac{1}{4\pi}\int_{\Omega}d\Omega\,W_{S}(\theta,\phi,\alpha,\beta) \boldsymbol{Y}^{m}_{l}(\theta,\phi).
\end{align}
One finds that each harmonic component is a product of two factors: the first piece depends only on time and gives the scale of GW whereas the second one depends solely on observer's position and can be obtained irrespective of details of the neutrino emission. The first factor can be evaluated once the angle-dependent neutrino luminosity $dL/d\Omega$ is provided by reliable multi-dimensional  simulations of PNS cooling, which are unfortunately still unavailable for the moment. In this paper, we hence employ some models for the evaluation of $h^{\rm amp}_{lm}$.

\subsection{Models}
\noindent In the formulation given above, the input is $a_{lm}(t)$, which can be calculated from the luminosity once it is provided by simulations. Unfortunately, we are currently lacking such simulations. In this paper, we will hence give $a_{lm}(t)$ by hand. In doing so, we take the simplest assumption: all the harmonic components of the luminosity, both axisymmetric and non-axisymmetric ones, have the same (but scaled) temporal profile as the isotropic one: $a_{lm}(t) = \epsilon a_{00}(t)$. This is certainly not true: each component has its own time evolution in reality. We believe, however, that this treatment still gives us an insight into the information we can retrieve from the GW signals. Note that the angular dependence of each harmonic component is fully taken into account in advance in this formulation. \\
We consider three models in this paper. In model~1a and~1b, we employ piecewise exponential functions (PEFs) for the temporal profile of the isotropic component (and hence all harmonic components) as follows:\\
model~1a:
\begin{align}
&a_{00}(t)= A_{i}e^{-t/t_{i}}\left[10^{53} \rm erg/s\right],\notag\\ 
&(A_{i},t_{i})=
\left\{
\begin{array}{lcl}
(1.20,0.008),  & & 0.000\leqslant t [\rm sec] < 0.004; \\
(0.55,0.500),  & &0.004\leqslant t [\rm sec] < 0.300 ; \\
(0.35,4.000),  & &0.300\leqslant t [\rm sec] < 3.600; \\
(0.15,12.50),  & &3.600\leqslant t [\rm sec] < 20.00; \\
(0.07,25.00),  & &20.00\leqslant t [\rm sec] \leqslant 50.00.
\end{array} \right.
\end{align}
model~1b:
\begin{align}
&a_{00}(t)= A_{i}e^{-t/t_{i}}\left[10^{53} \rm erg/s\right],\notag\\  
&(A_{i},t_{i})=
\left\{
\begin{array}{lcl}
(3.000,0.008), & &    0.000\leqslant t [\rm sec] < 0.004; \\
(1.800,0.500),  & &   0.004\leqslant t [\rm sec] < 0.300; \\
(1.100,2.000),  & &   0.300\leqslant t [\rm sec] < 3.600; \\
(0.400,5.000),  & &   3.600\leqslant t [\rm sec] < 20.00; \\
(0.018,20.00),  & &   20.00\leqslant t [\rm sec] \leqslant 50.00. 
\end{array} \right.
\end{align}
Model 1a roughly fits the result of a 1D simulation of PNS cooling, which itself is employed as model~2. Model 1b has shorter cooling times for the reason explained later. All these models both displayed in Fig.~2. \\
In model 3, we assume that the neutrino emission is axisymmetric with respect to a certain axis denoted by the $\it Z''$-axis; this axis is rotating at a constant angular frequency $\omega$ around the $\it Z$-axis of the space-fixed coordinates ($\it X$,~$\it Y$,~$\it Z$) (see Fig.~3).
Then the neutrino emission is non-axisymmetric with respect to this space-fixed coordinates in a time-dependent way.\\
We can derive this time-dependent non-axisymmetric distribution of neutrinos by considering the transformation from the rotating coordinates ($\it X''$,~$\it Y''$,~$\it Z''$), in which the neutrinos are axisymmetrically emitted with respect to $\it Z''$-axis, to the space-fixed coordinates ($\it X$, $\it Y$, $\it Z$). This is done as follows: 
in Eq.~(10), the sum of  $\boldsymbol{Y}^{m}_{l}(\theta,\phi)$ is replaced by $P_{l}(\cos\theta')$, where $\theta'$ is the angle between the $\it Z''$-axis and the flight direction of neutrino direction $n'$ (see Fig.~3); we use the notation $b_{l}(t)=\epsilon a_{00}(t)$ for convenience; we then apply the addition theorem to the $P_{l}(\cos\theta')$ as follows:
\begin{align}
    &\frac{dL(t,\theta,\phi)}{d\Omega}=
    \sum^{\infty}_{l=0}\frac{b_{l}(t)}{4\pi}P_{l}(\cos\theta')\notag\\
     &=\sum_{l=0}^{\infty}\sum_{m=-l}^{l}b_{l}(t)\frac{4\pi}{2l+1}
     \boldsymbol Y_{l}^{m}(\Theta(t),\Phi(t))\left[\frac{1}{4\pi}\boldsymbol Y_{l}^{m}(\theta,\phi)\right],
\end{align}
with
\begin{align}
    &\Theta (t)=\mbox{const.,}\\
    &\Phi (t)=\omega t,
\end{align}
where $\omega$ is the angular frequency of rotation. Combining $b_{l}(t)$ and $\boldsymbol{Y}^{m}_{l}(\Theta(t), \Phi(t))$ to give $a_{lm}(t)$ as
\begin{equation}
    a_{lm}(t)=b_{l}(t)\boldsymbol{Y}^{m}_{l}(\Theta(t),\Phi(t)),
\end{equation}
which leads to
\begin{align}
    h^{\rm amp}_{lm}(t)=\frac{8\pi G}{c^{4}R(2l+1)}\int^{t}_{0}a_{lm}(t')dt',
\end{align}
we finally get
\begin{align}
    h_{S}(t,\alpha,\beta)=\sum^{+\infty}_{l=1}\sum^{l}_{m=-l}h_{lm}^{\rm amp}(t)\times\Psi_{lm}^{S}(\alpha,\beta),
\end{align}
the same form as in Eq. (11).
\begin{figure}
\centering 
\includegraphics[width=0.5\textwidth]{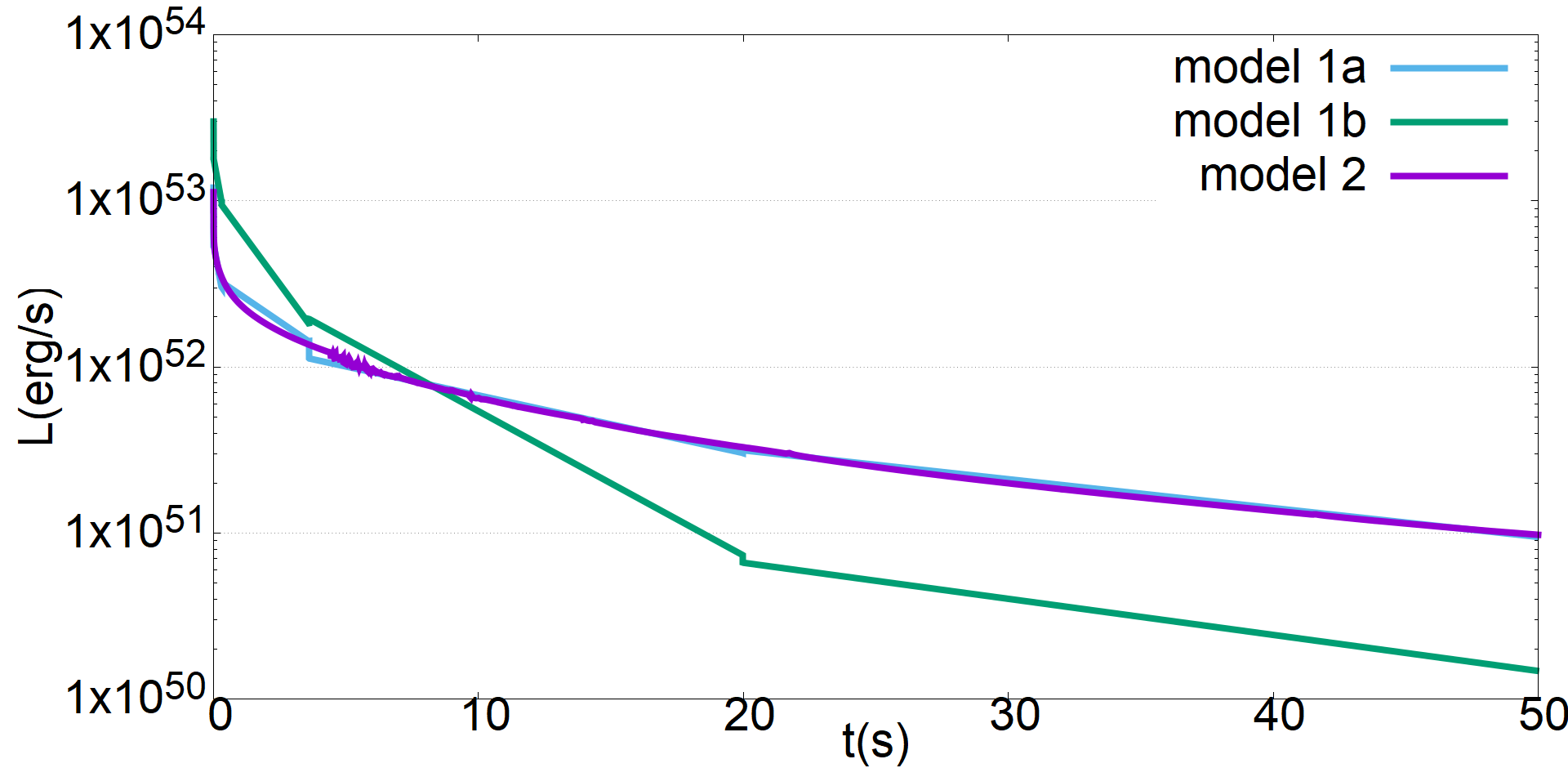}
\caption{The isotropic component of neutrino luminosity as a function of time. The purple line is the result of a 1D simulation of PNS cooling and used in model 2. The green and blue lines are the PEFs employed for models 1a and 1b, respectively.}
\end{figure}
\begin{figure}
\centering 
\includegraphics[width=0.5\textwidth]{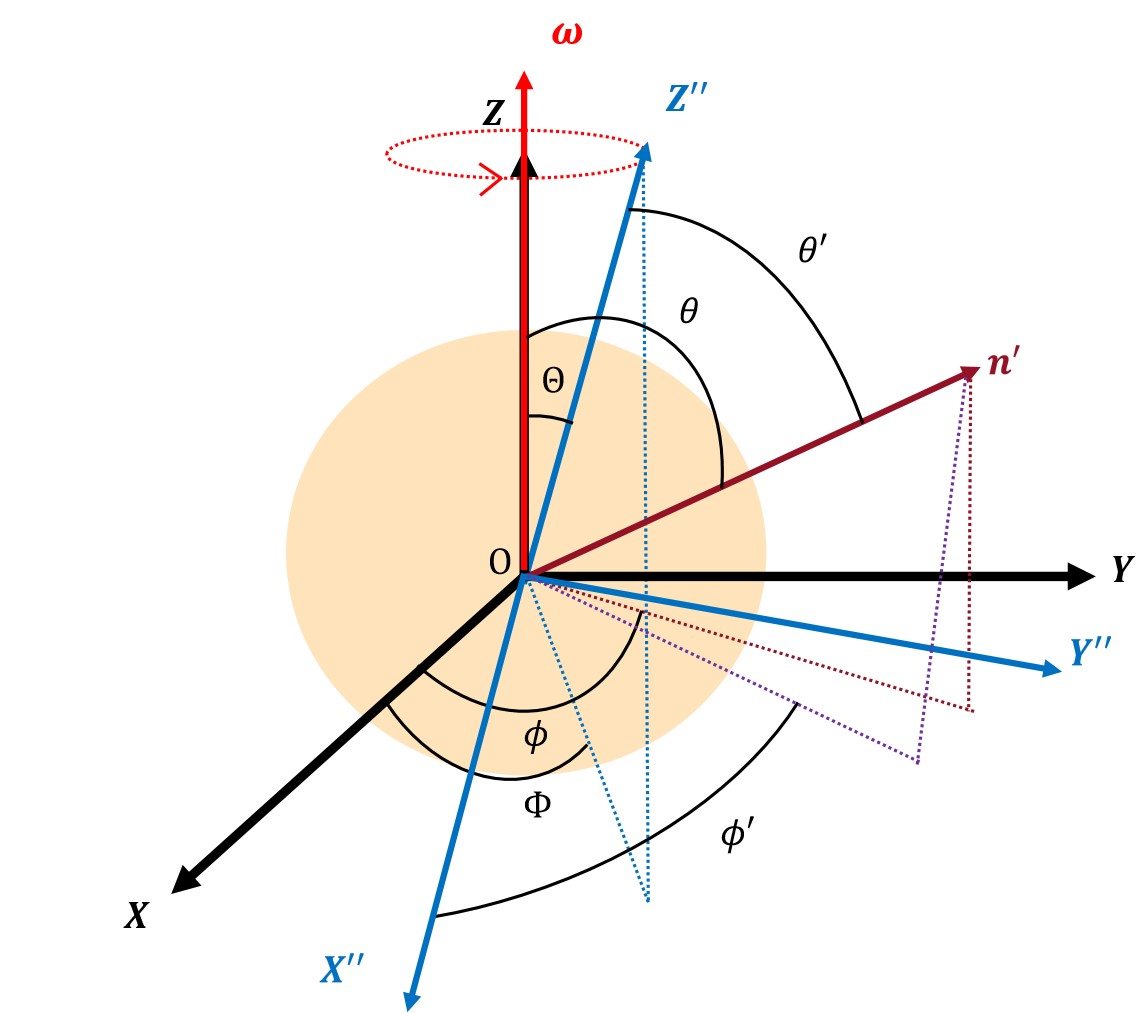}
\caption{The space-fixed coordinates ($\it X$, $\it Y$, $\it Z$) and the co-rotating coordinates ($\it X''$, $\it Y''$, $\it Z''$) employed in model 3.}
\end{figure}

\noindent Now we explain the ideas behind these models somewhat in detail. In the current situation, in which no reliable multi-dimensional simulation of PNS cooling is available, the results of 1D simulations under spherical symmetry are the best information we can utilize. In model 2, we adopt indeed the time profile obtained in one of such 1D simulations for the isotropic component of the neutrino luminosity (and hence for all harmonic components after scaled) in our calculation of GW signals. It is well known, on the other hand, that the PNS cooling has a few phases that have different cooling timescales\cite{Pons_1999}. Considering the experimental nature of our study in this paper, we approximate the light curve with piece-wise exponential functions in models 1a and 1b so that we could see the different cooling times more clearly. Model 1a is indeed a rough fit to the original light curve. In model 1b, on the other hand, the cooling times in phases 3 through 5 are reduced from the values adopted in model 1a.~This is motivated by the expectation that convection will occur in PNS and accelerate the cooling by enhancing neutrino emissions. Incidentally, in these experimental models we investigate separately the time dependence of the GW signals and the angular dependence of each harmonic component, which is possible with our formulation (see Eq. (11)).\\
In model~2 as we mentioned, we employ the result of a 1D simulation of PNS as it is. Although the detail of this simulation is not important as long as it is of the standard quality, we will give some information briefly: the initial condition is the snapshot at t=0.3s post bounce obtained in a 1D core-collapse simulation for $15M_{\bigodot}$ progenitor\cite{1995ApJS101181W} with the general relativistic neutrino-radiation hydrodynamics code\cite{Sumiyoshi_2005}; the inner region of the core (up to $\sim 1.47M_{\bigodot}$) is extracted; the subsequent quasi-static evolution of PNS cooling is computed by solving the Tolman-Oppenheimer-Volkov equation together with the neutrino transfer equation under the so-called flux-limited diffusion approximation\cite{Suzuki_1990vn}; Shen's equation of state is used\cite{2011ApJS19720S}. Let us stress again that the employment of the 1D-simulation result is a crude approximation. Even the isotropic component will be different in reality. We need to wait for multi-dimensional simulations of PNS cooling.\\
Model 3 is more physically motivated. In this model neutrinos are emitted axisymmetrically with respect to an axis that is rotating at a constant angular frequency around another axis fixed in space. Such a situation may occur for a rotating magnetized PNS with the magnetic axis misaligned with the rotation axis, in which neutrinos may be emitted more preferentially in the direction of the magnetic axis\cite{Kuroda_2021}. Since we are concerned with the GW emission in the deci-Hz range in this paper, slow rotators are the target. Magnetars are known to rotate slowly with the rotation period longer than a second typically although we do not know if they are born as such slow rotators. The very strong magnetic fields $B\gtrsim 10^{14}$G that characterize the magnetars are also advantageous for this scenario. Electrons will be affected by such strong fields and in turn influence the neutrino transport\cite{Kuroda_2021}, picking up the magnetic axis as a special direction. Although the observational information on the magnetic axis in magnetars is scarce, it is more natural that the two axes are not aligned with each other. This statement may be supported by the fact that
pulsars are supposed to be misaligned rotators\cite{2004hpa..book.....L}\cite{1968Natur.217..709H}\cite{Buckley_2017}\cite{Tan_2018}. There are also some 3D magnetohydrodynamic (MHD) simulations that suggest the misalignment of the rotational and magnetic axes\cite{02148}\cite{Bugli_2021}.\\
In the actual evaluation of GW signals, the scaling factor for the anisotropic components of the luminosity is set to $\epsilon =0.1$ in models 1 and 2. For model 3, on the other hand, we choose $\epsilon$ = 0.01, which may correspond to $\it B$ $\sim 10^{15}$G\cite{Kuroda_2021}. We have two more parameters to set in this case: $\Theta$ and  $\omega$.
Since we focus on the deci-hertz range, we choose the spin frequency as $f= \omega/2\pi = 0.1\rm Hz$. The inclination angle $\Theta$ is arbitrary and chosen to be $\Theta=\pi/2$ as an example. Note that different choices of $\Theta$ change the amplitude alone (see Eqs.~(19) and (20)).

\section{\label{sec:level1}Results}
\noindent Now we present the main results. Taking advantage of the fact that each harmonic component of the GW waveform (and hence the characteristic strain also) is decomposed into the two factors (see Eq.~(11)): $h_{lm}^{\rm amp}$ and $\Psi_{lm}^{S}$, each depending on the time and the observer's position alone, respectively, we will investigate them separately. We first show the waveforms $h_{lm}^{\rm amp}(t)$ and the corresponding characteristic strains for models 1 and 2 in subsections A and B, respectively. We then look at the anisotropies of the GW emissions based on $\Psi_{lm}^{S}(\alpha, \beta)$. We also present the polarization angles of each harmonic mode up to $l = 3$. The results of model 3 will be given thereafter. Again the waveform, characteristic strain as well as the polarization will be studied in detail. Finally, we discuss prospective detections in the deci-hertz range by some planned satellite-borne detectors.
\subsection{$h_{lm}^{\rm amp}(t)$ and the characteristic strains for models 1 and 2} \label{KeepUp}
\noindent In this section, we present $h_{lm}^{\rm amp}(t)$ up to 50s, and the corresponding characteristic strains $\tilde{h}_{lm}^{c}(f)$ for models 1 and 2. Note that $h_{lm}^{\rm amp}(t)$ gives the overall scale and the time profile of GW signals, whereas $\Psi_{lm}^{S}(\alpha, \beta)$ gives the angular dependence of the GW emissions. We evaluate Eq.~(12) numerically. As mentioned repeatedly, the time profiles $h^{\rm amp}_{lm}(t)$ are assumed to be the same for all harmonic components in models 1 and 2. In model 2, in particular, we use the time profile obtained in the 1D simulation of PNS cooling just by scaling it. Then Eq.~(12) is reduced to 
\begin{equation}
    h_{lm}^{\rm amp}(t)=\frac{2G}{c^{4}R}\int^{t}_{0}\epsilon L_{\nu}^{1D}(t')dt',
\end{equation}
where $L^{1D}_{\nu}(t)$ is the luminosity obtained in the simulation; the scaling factor $\epsilon$ is set to 0.1. In models 1a and 1b, on the other hand, $L^{1D}_{\nu}(t)$ is replaced with the PEFs given in Eqs.~(14) and (15), respectively. These two models are of experimental nature as can be understood from Fig.~2. \\
The GW waveforms obtained for these models are shown in Fig.~4. Although the approximation by the PEFs is rather crude, there is a good agreement between models~1a and 2. Note that there are slight differences in the asymptotic values among the different models, since we did not make any attempt to make the values of the integral in Eq. (22) for these two models close to that for model 2 when we constructed models 1a and 1b. The waveform is actually very simple, increasing monotonically. It is also clear that the GW amplitudes approach nonvanishing asymptotic values. This is the memory effect and reflects the fact that neutrinos continue to expand as the time goes to infinity, in sharp contrast to matter motions, which are eventually halted. For model~1b the GW amplitude rises more quickly in accord with the faster decline of the neutrino luminosity. We are not concerned with a slight difference in the asymptotic values among models as we mentioned earlier.
\begin{figure}
\centering 
\includegraphics[width=0.5\textwidth]{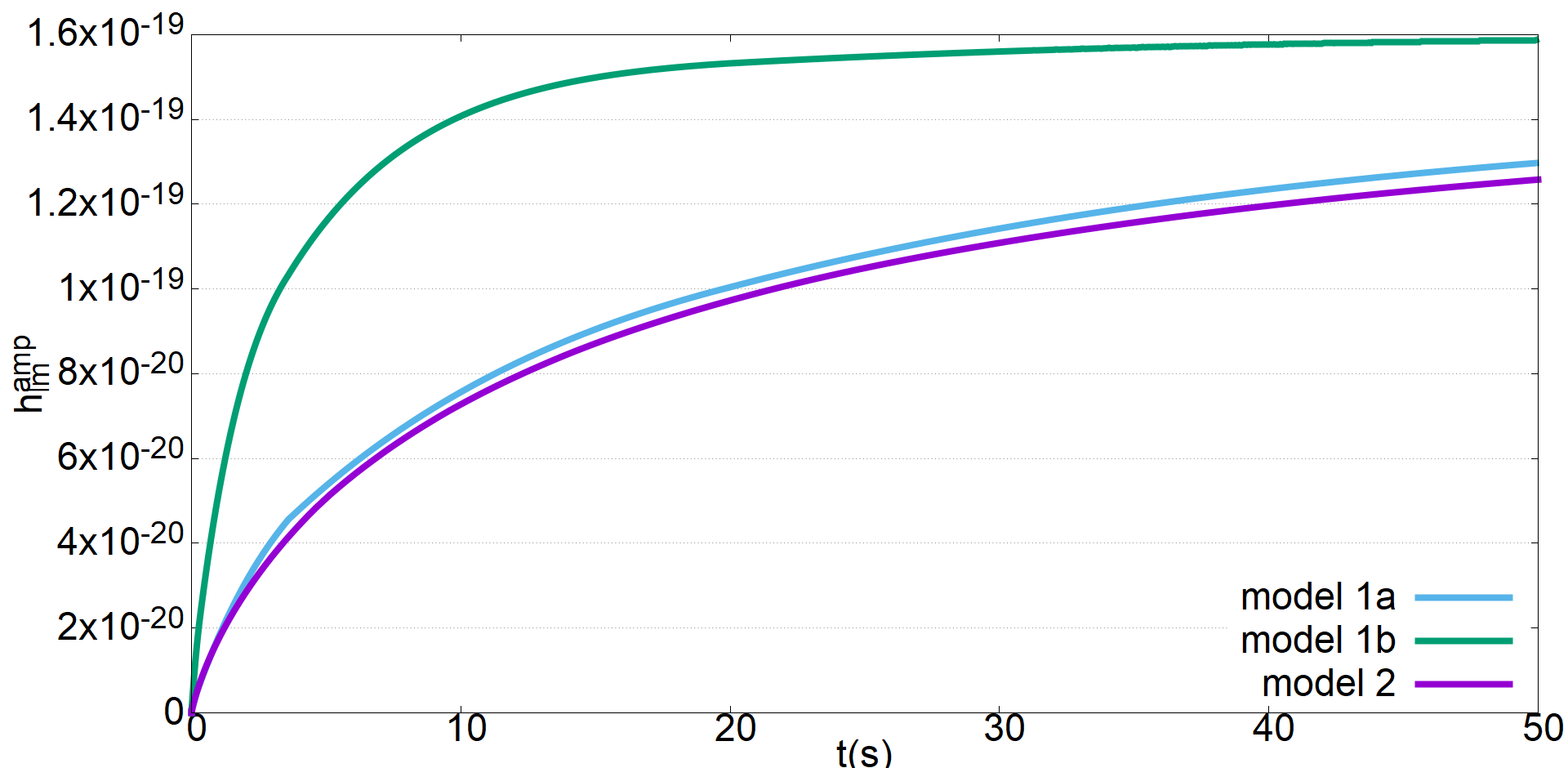}
\caption{The gravitational waveforms for models 1a, b and 2. The angular factor $\Psi^{S}_{lm} (\alpha,\beta)$ is not included.}
\end{figure}

\noindent The characteristic strain $\tilde{h}_{S}^{c}(f)$ of GW is a dimensionless quantity, which is commonly used for the comparison of a signal to a detector sensitivity. It is defined as 
\begin{equation}
    \tilde{h}_{S}^{c}(f)=2f|\tilde{h}_{S}(f)|,
\end{equation}
where $\tilde{h}_{S}(f)$ is the Fourier transform of the waveform $h_{S}(t)$;
\begin{equation}
    \tilde{h}_{S}(f)=\int^{\infty}_{-\infty}h_{S}(t)e^{2\pi ift} dt.
\end{equation}
The characteristic strains for the PEFs can be calculated analytically as follows.\\
Each exponential function in Eqs.~(14) and (15) is written in its own range as 
\begin{equation}
    L_{\nu i}(t)=k_{i}e^{-\frac{t}{t_{i}}},\quad i=1,2,3,4,5,
\end{equation}
and the real and imaginary parts of its Fourier transform are given, respectively, as
\begin{align}
     {\rm Re} \left(\tilde{L}_{\nu i}(\omega)\right)=&\frac{-t_{i}}{1+(\omega t_{i})^{2}}[
    e^{-\frac{T_{i}}{t_{i}}}(\cos\omega T_{i}-\omega t_{i}\sin\omega T_{i})+\notag\\
    &e^{-\frac{T_{i-1}}{t_{i}}}(\omega t_{i}\sin\omega T_{i-1}-\cos\omega T_{i-1})
    ],
\end{align}
\begin{align}
    {\rm Im} \left(\tilde{L}_{\nu i}(\omega)\right)=&\frac{t_{i}}{1+(\omega t_{i})^{2}}[
    e^{-\frac{T_{i}}{t_{i}}}(\omega t_{i}\cos\omega T_{i}+\sin\omega T_{i})-\notag\\
    &e^{-\frac{T_{i-1}}{t_{i}}}(\omega t_{i}\cos\omega T_{i-1}+\sin\omega T_{i-1})
    ],
\end{align}

\begin{figure}
\centering 
\includegraphics[width=0.5\textwidth]{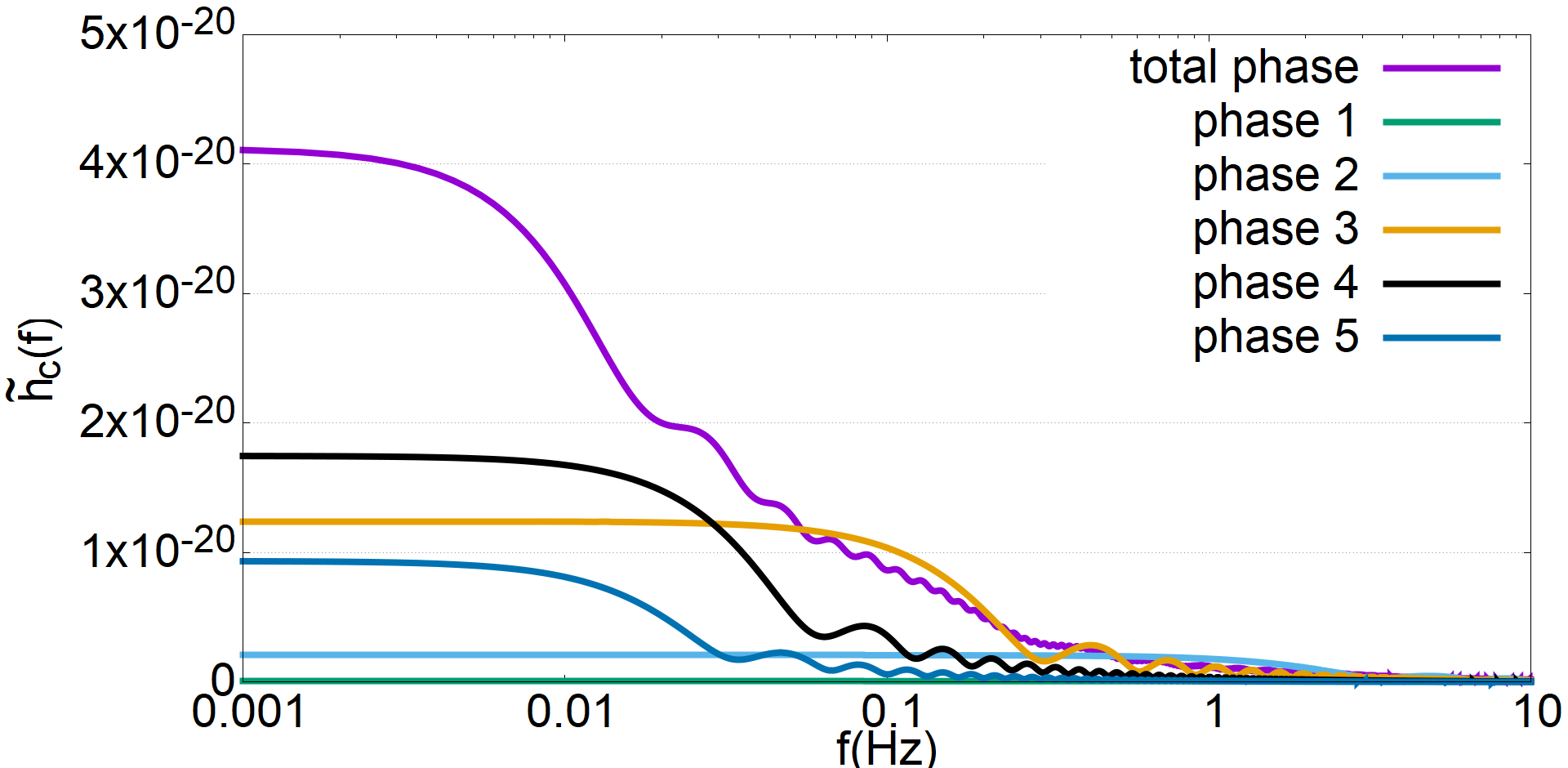}
\caption{The characteristic strain $\tilde{h}_{c}(f)$ for model 1a (purple line). Other lines with different colors are the contributions from the individual phases.}
\end{figure}

\begin{figure}
\centering 
\includegraphics[width=0.5\textwidth]{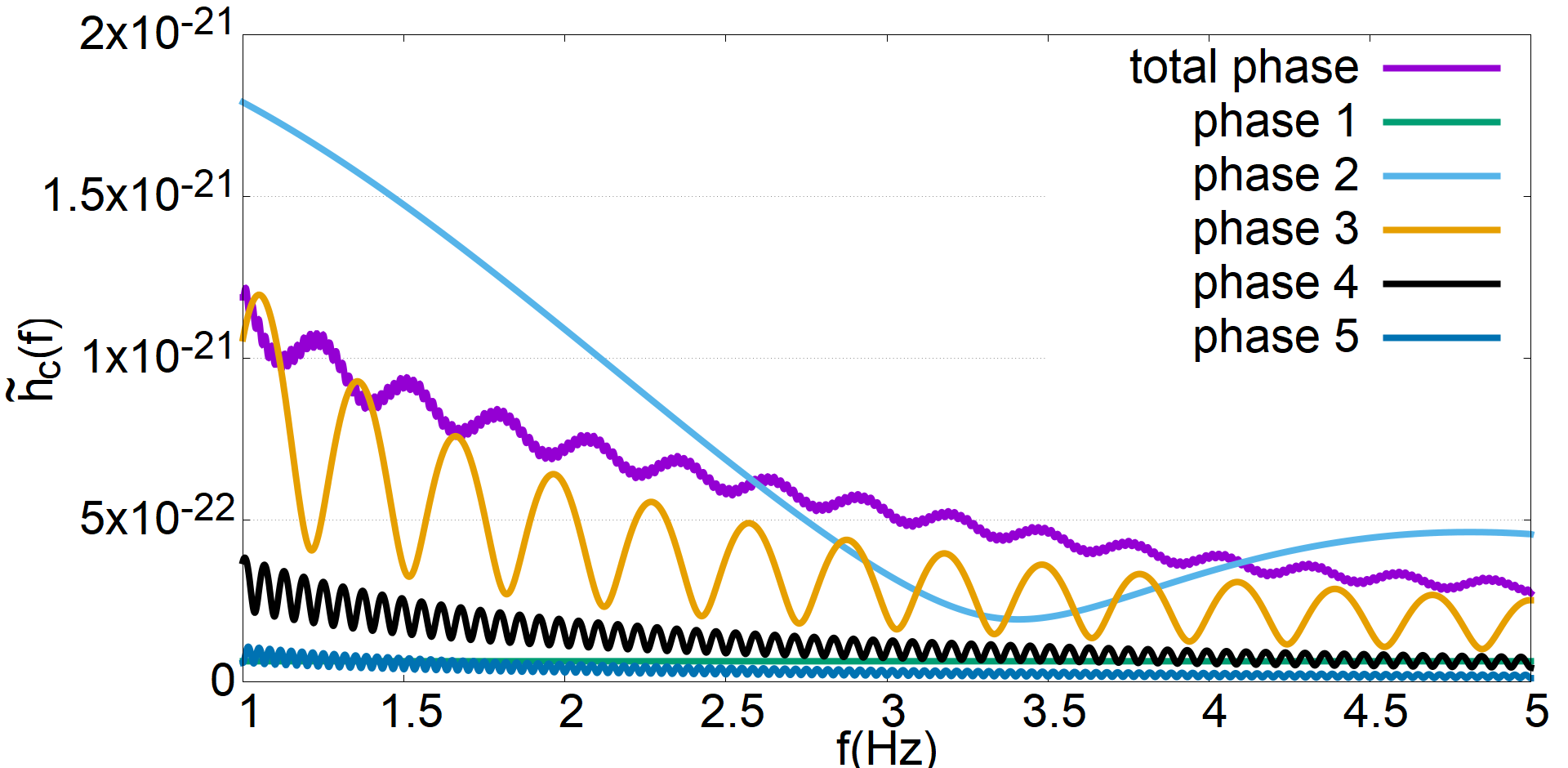}
\caption{The zoom-in to the frequency range of 1-5Hz of Fig.~5. The notations are the same as in Fig.~5 but the horizontal axis is in linear scale.}
\end{figure}

\noindent where $\omega=2\pi f$ and $T_{i-1}$ and $T_{i}$ are the lower and upper bounds of the i-th range. The Fourier transform of the total PEF is just the sum of the contributions from the individual ranges. \\
The characteristic strains so obtained are shown in Fig.~5. In this paper, as mentioned repeatedly, we will focus on the sub-hertz range. In this figure we show the contributions from individual phases separately in addition to the characteristic strain for the entire evolution. In the latter (the purple line) one can see clear features which we want to exploit to find the decay time of each phase. As a matter of fact, if one pays attention to the frequency range of 0.001-0.02Hz, one finds that the characteristic strain is almost constant up to $\sim$0.002Hz, which is nothing but the memory. 
Note that we cannot obtain the strictly zero-frequency limit of the characteristic strain from data that spans a finite period. Since the characteristic strain is a continuous function of frequency, however, it is expected to be nonvanishing and almost constant near the zero-frequency. Then the characteristic strain starts to decline and reaches another (short) flat region at $\sim$ 0.015Hz.
Comparing the purple line with the blue line, which is the contribution from phase 5, one recognizes that the decline in the former reflects that of the latter. It is also found that the frequency at the second plateau corresponds to twice the inverse of the decay time, $t_{5}=25$ sec, for phase 5.
Similarly, there is another decline and the subsequent flattening of the characteristic strain in the frequency range of 0.02-0.04Hz. The decline is again produced mainly by the fall in the contribution from phase 4. We find also that the frequency at the third plateau corresponds to twice the inverse of the decay time for the same phase, $t_{4}=12.5$ sec. \\
It is difficult, however, to extend this scheme to phase 3 or even earlier phases, since there are many oscillations up to the frequencies, at which the decline occurs in these early phases. There may be another way to extract the decay times for these phases, though, if one looks at a bit higher frequencies. 
Fig.~6 is a zoom-in to the frequency range of 1 - 5Hz. One can see the characteristic strain oscillating in this range. It is also apparent from the comparison with the contribution from phase 3 (orange line) that this reflects the oscillation of the characteristic strain in this phase. One finds then that the number of oscillations per frequency, $\sim$3.7, corresponds to the decay time of the same phase. If one inspects the purple line, i.e., the characteristic strain for the entire evolution more closely, one finds that this oscillation is actually superimposed by other oscillations with smaller amplitudes and shorter periods, which reflect in turn the oscillations in phases 2 and 1. In principle, the same method can be applied to the first and second phases by looking at even higher-frequency ranges, where the oscillations are dominated by the contributions from these early phases. In practice, however, this will be difficult to do, since the amplitudes are much smaller in fact and, moreover, at these higher frequencies we are afraid that GWs from matter motions, which we do not consider here, may not be ignored.\\
\noindent We now shift our attention to model 1b, which has shorter decay times in phases 3 to 5 and may correspond to the accelerated PNS cooling in the presence of convection\cite{2006}\cite{2012}. We present the characteristic strain for the entire evolution as well as the individual contributions from different phases in Fig.~7. One finds from this figure that the characteristic strain (purple line) is much feature-less with clear humps and dips being absent. It is hence difficult to apply the previous methods to find the decay times. Comparing the characteristic strain for the whole evolution (purple line) with the contributions from the individual phases (lines with other colors), however, one still finds that the decay times for these phases are encoded as the changes of slope in the characteristic strain. In fact, the first change of slope, i.e., the one at the lowest frequency, occurs at $\sim$ 0.03Hz, which should correspond to a decay time $\approx \frac{1}{2\times 0.03}=16.7 \rm s$. This is indeed close to the decay time of phase 5, $t_{5} = 20 \rm sec$, in Eq.~(15). Similarly, the second change of slope, although tiny, can be found at the frequency of $\sim$ 0.1Hz, which actually gives the decay time of phase 4, i.e.,  $t_{4}=5\rm s$. 
Unfortunately, the decay time of phase 3 is almost impossible to find although it is the largest contribution in this model. This is because the sign of the contribution from phase 3, which is not shown in the figure, is opposite to those of the contributions from other phases and its change of slope is almost washed out. Since the extraction of the decay times for these late phases are already very subtle admittedly, we did not make any attempt to retrieve the decay times of the earlier phases. Note that we determine the above frequencies for the plateaus and the points of slope by eye inspection, since we think that it is sufficient for our purpose in this paper.\\
In Fig.~8, we present the characteristic strains for model~2, that is, for the time profile obtained in the 1D PNS cooling simulation. For comparison, we show the results of models 1a and 1b with the PEFs given in Eqs.~(14) and (15), respectively. One finds that the characteristic strain for model 2 is almost identical to that for model 1a. This implies that the decay times at different evolutionary phases can be extracted also from the realistic time profile just in the same way.\\ 
Note in passing that the common feature observed in Fig.~8 that the characteristic strain has a non-vanishing zero frequency limit (ZFL) is a manifestation of the GW memory. In fact, the characteristic strain is defined as 
\begin{equation}
    \tilde{h}_{c}(f)=\left|\frac{i}{\pi}\tilde{\dot{h}}\right|
\end{equation}
and its ZFL can be expressed as
\begin{align}
    \mathop{\lim} \limits_{f\to 0} \tilde{{h}}_{c}(f)&=
    \mathop{\lim} \limits_{f\to 0} \left|\int ^{+\infty}_{-\infty}\frac{i}{\pi}\dot{h}(t)e^{i2\pi ft}dt\right|\notag\\
    &=\notag
    \left|\int ^{+\infty}_{-\infty}\frac{i}{\pi}\dot{h}(t)dt\right| =\frac{|h(t=\infty)|}{\pi},\\
\end{align}
where we assume that the $h(t=-\infty)=0$.\\
\begin{figure}
\centering 
\includegraphics[width=0.5\textwidth]{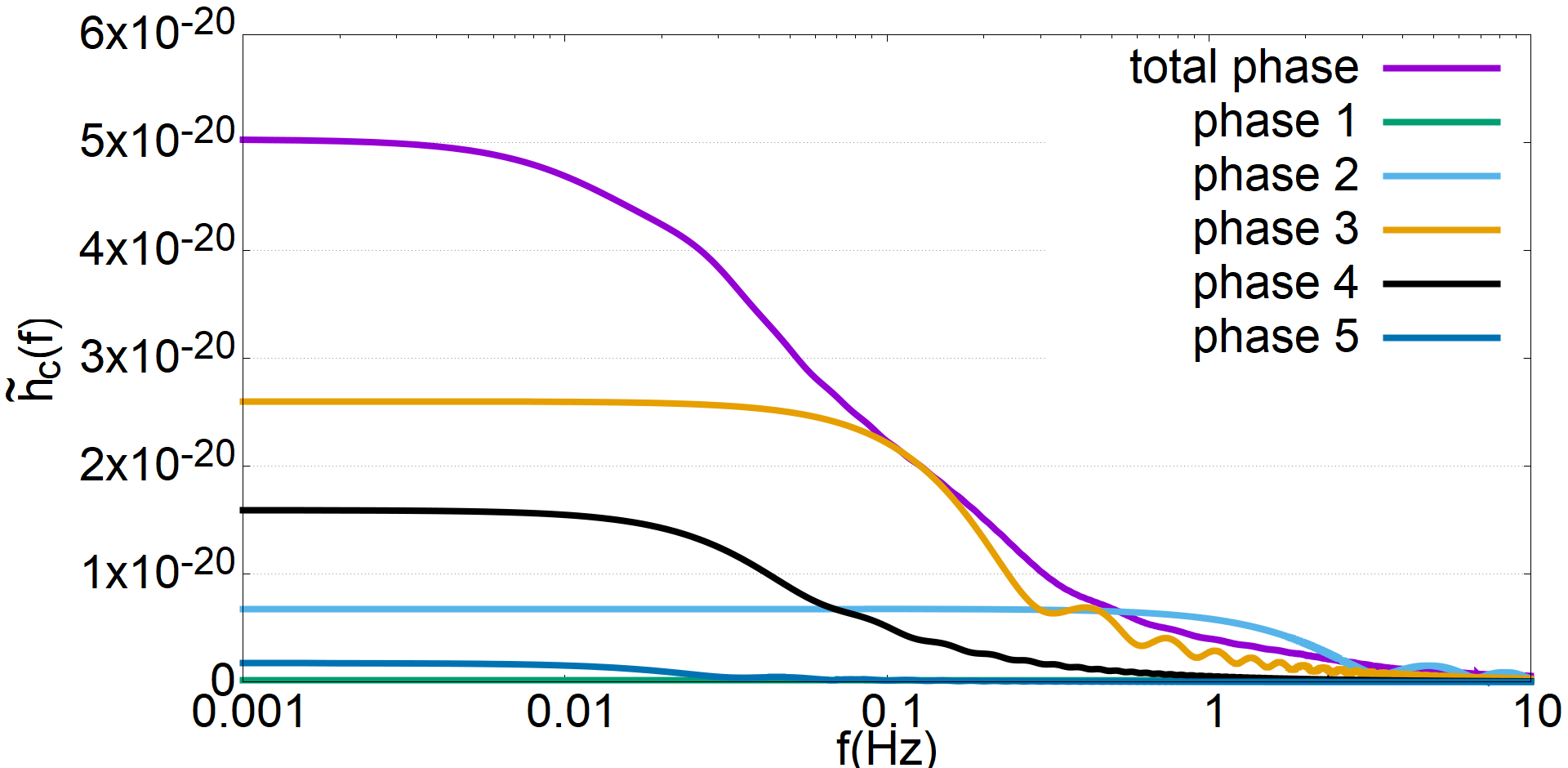}
\caption{The characteristic strains for model 1b. The notations are the same as in Fig.~5}
\end{figure}
\begin{figure}
\centering 
\includegraphics[width=0.5\textwidth]{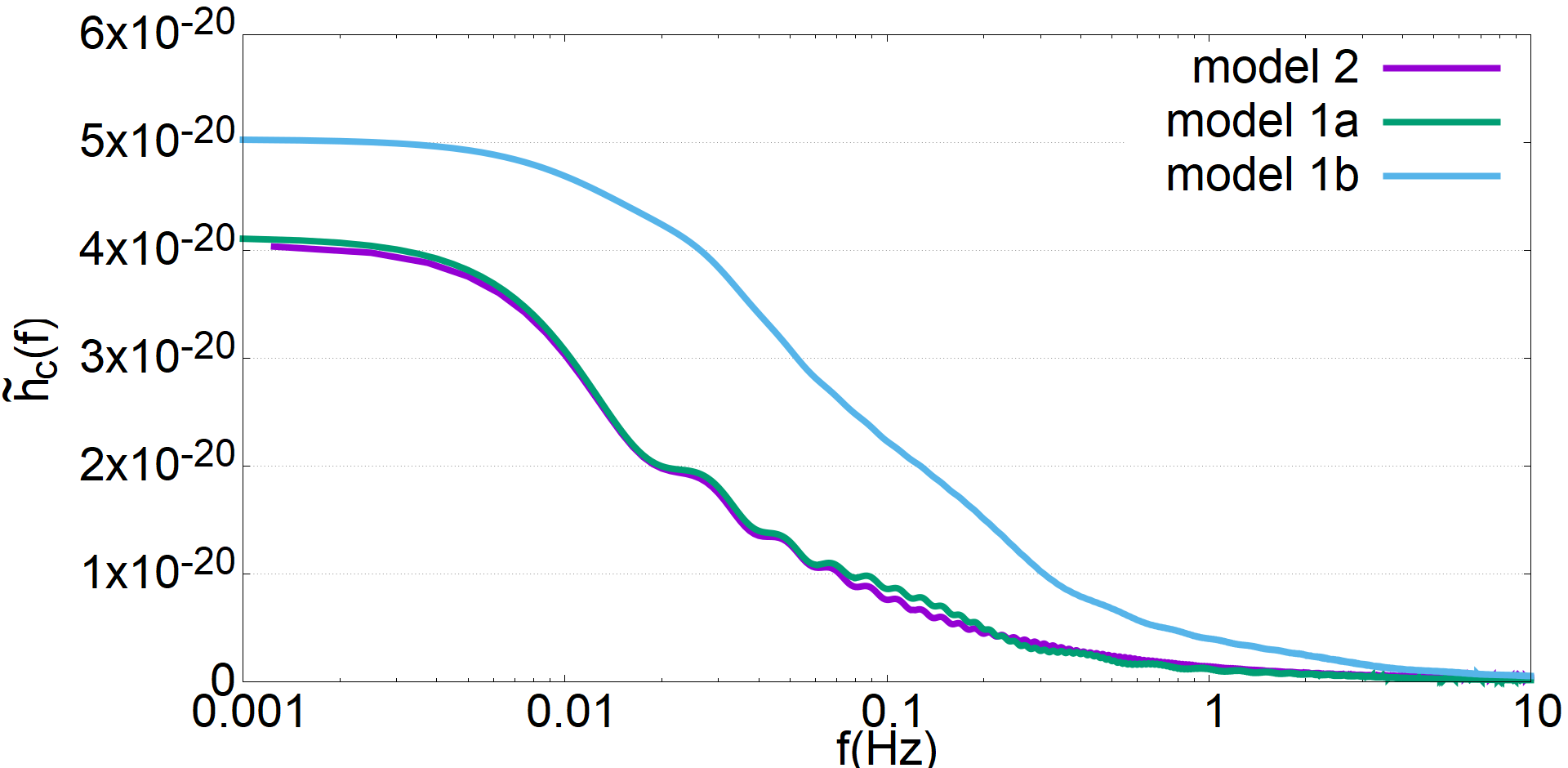}
\caption{The comparison of characteristic strains among models 1a, 1b and 2.}
\end{figure}
\begin{figure}
\centering 
\includegraphics[width=0.5\textwidth]{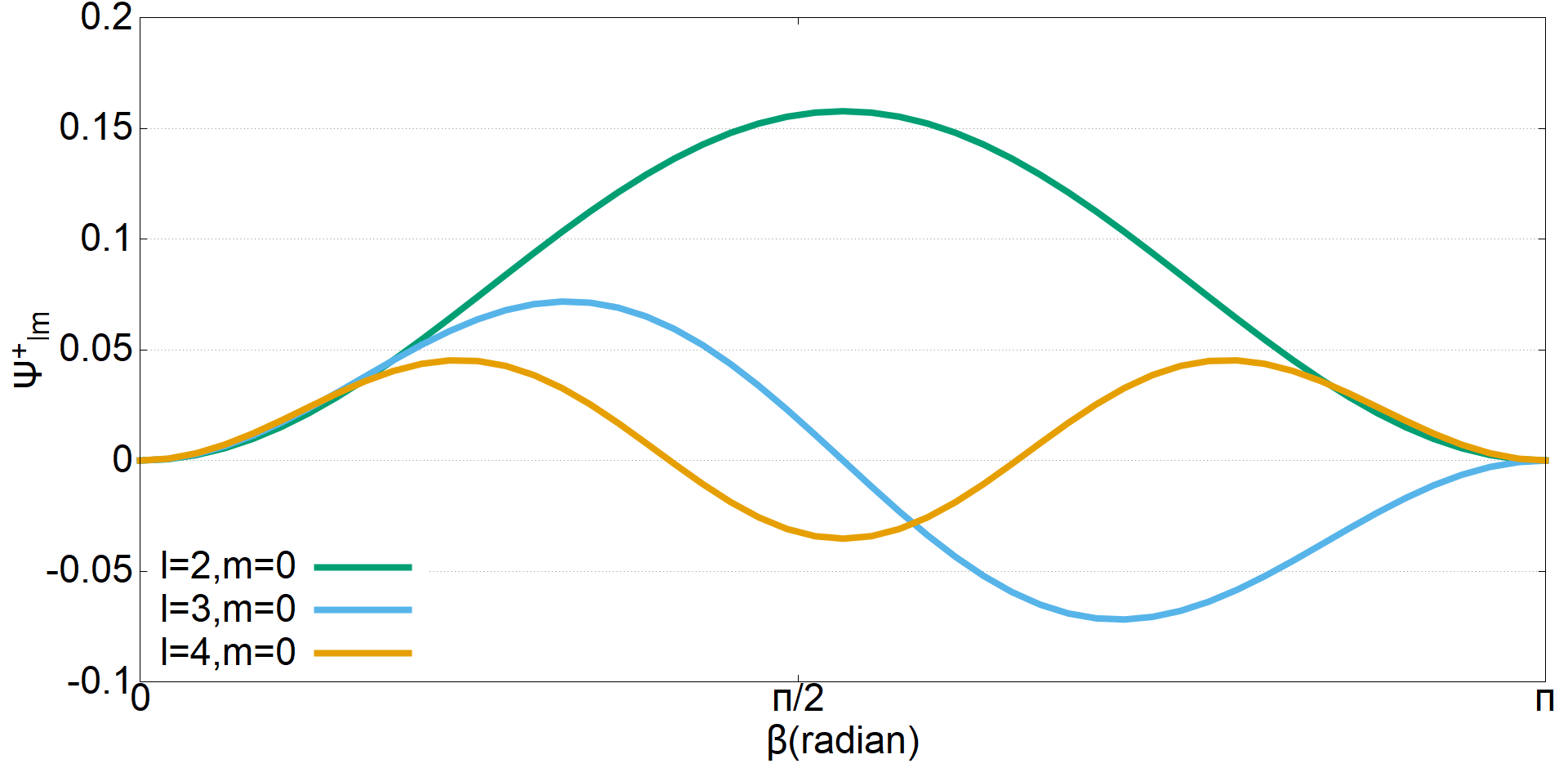}
\caption{The angular factor $\Psi^{+}_{lm}$ as a function of $\beta$. Here we set $\alpha=0$. Note that the $\times$-mode vanishes identically.}
\end{figure}
\begin{figure*}
\includegraphics[width=1\textwidth]{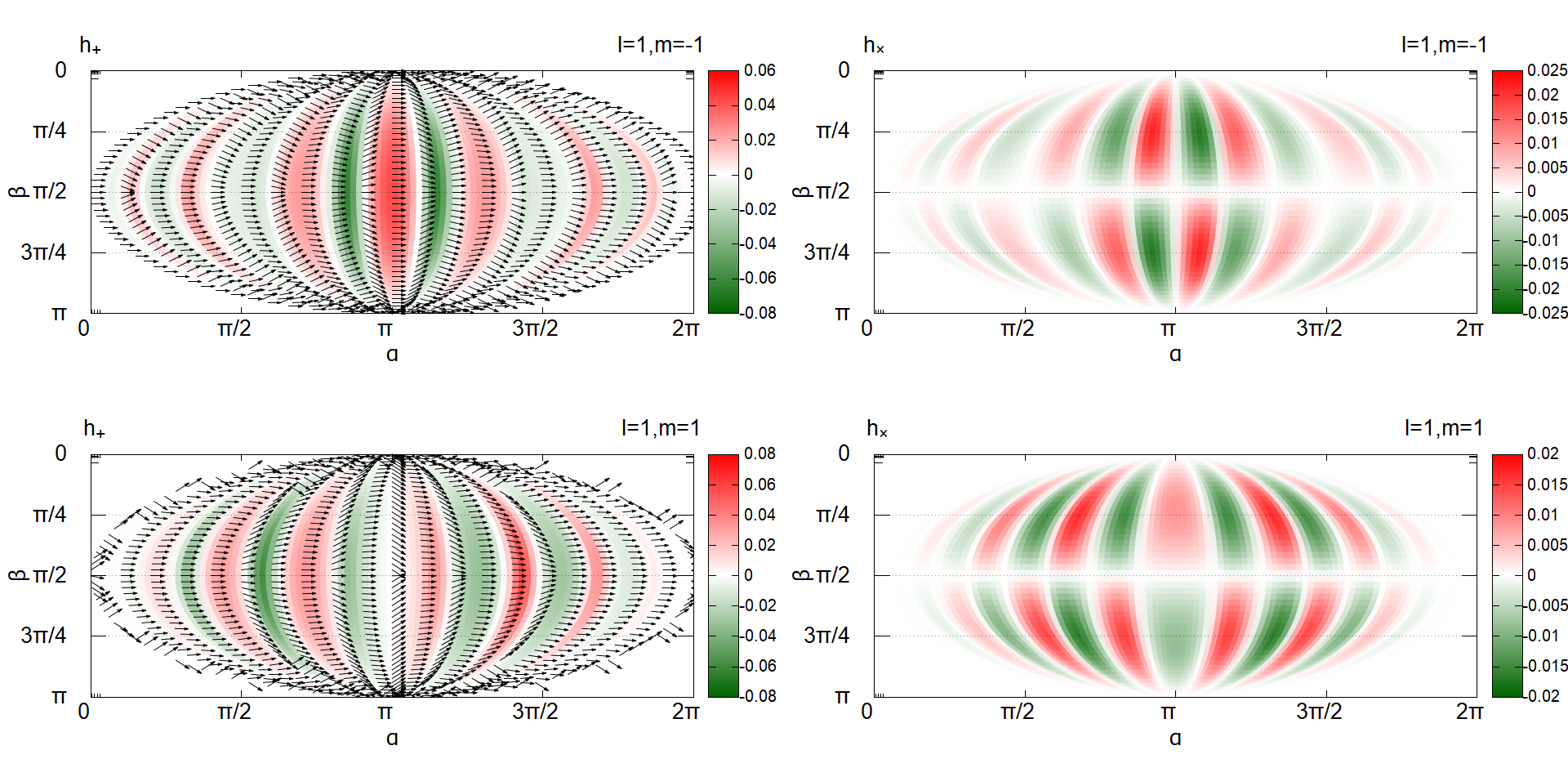}
\caption{\label{fig:wide}The color map of the angular factor $\Psi_{lm}^{S}(\alpha, \beta)$ for $l = 1$ and $m = \pm 1$ in the Mollweide projection. The left and right columns correspond to the +- and $\times$-modes, respectively. The arrows in the left panels represent the angles of the linear polarization for each harmonic component. The top and bottom rows are for the $m = - 1$ and $m = 1$, respectively. Note that the color scale is different from panel to panel.}
\end{figure*}

\begin{figure*}
\includegraphics[width=1\textwidth]{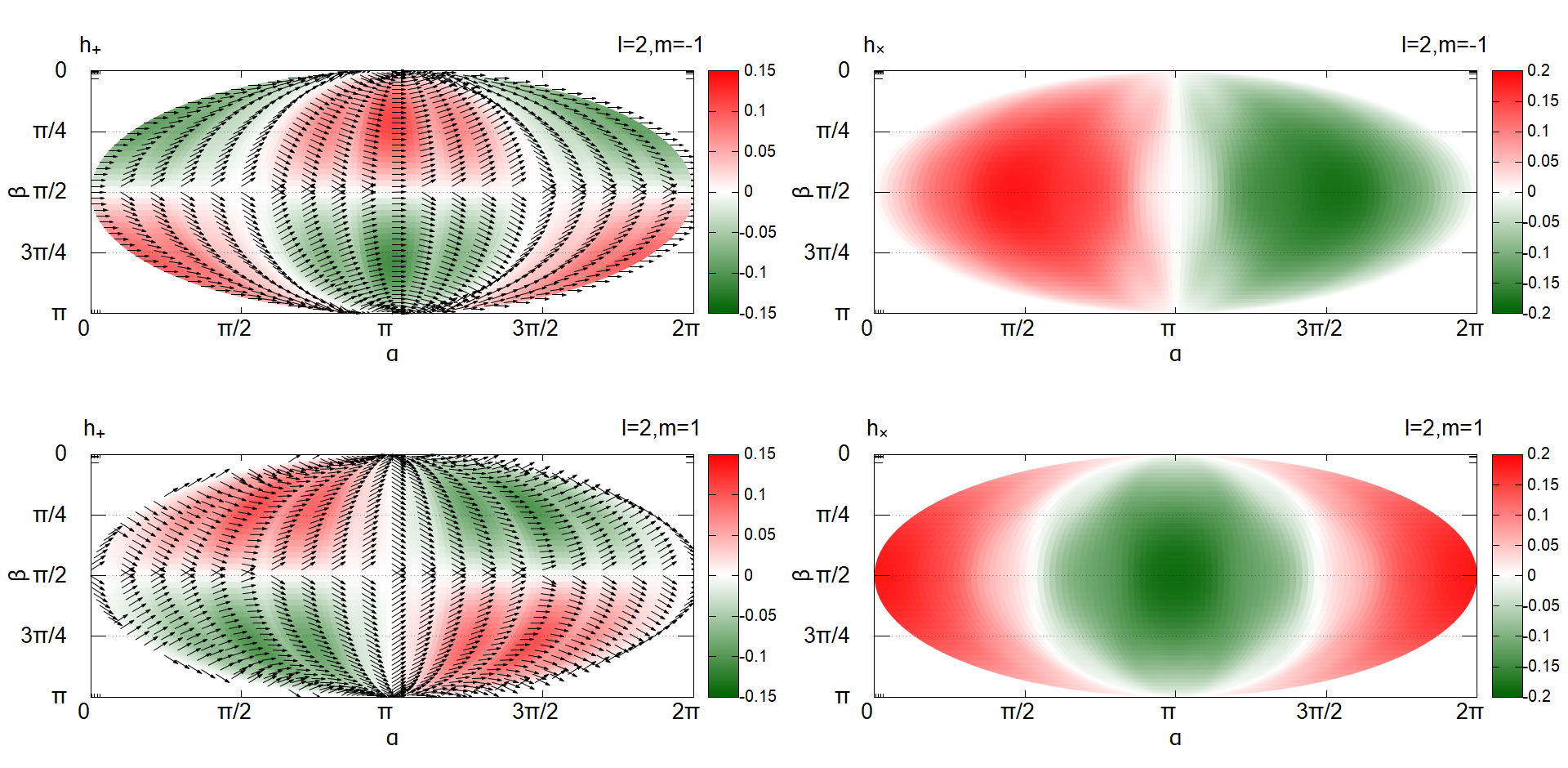}
\caption{\label{fig:wide}Same as Fig.~10 but for $l=2$ and $m=\pm 1$.}
\end{figure*}

\begin{figure*}
\includegraphics[width=1\textwidth]{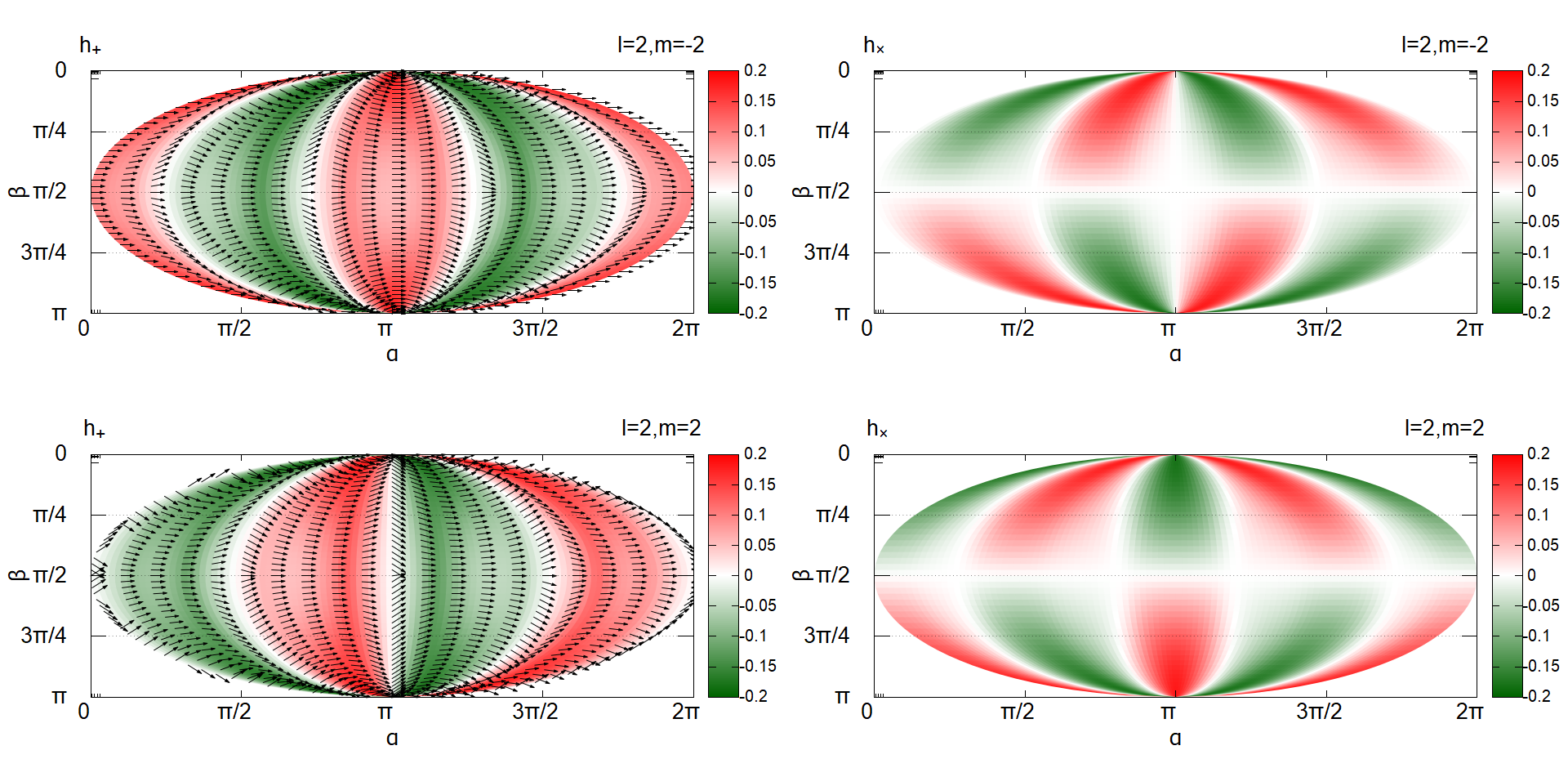}
\caption{\label{fig:wide}Same as Fig.~10 but for $l=2$ and $m=\pm 2$.}
\end{figure*}

\begin{figure*}
\includegraphics[width=1\textwidth]{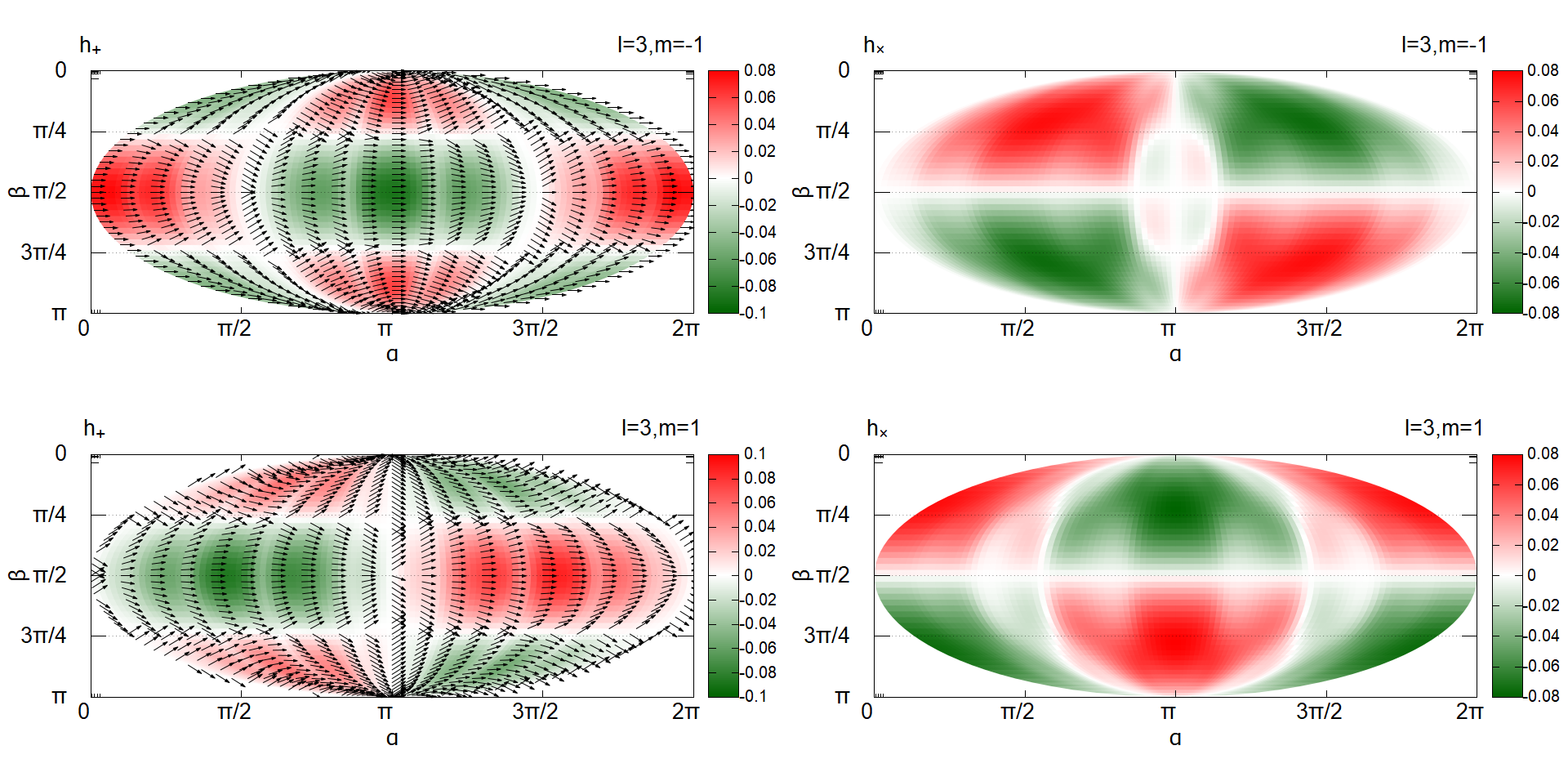}
\caption{\label{fig:wide}Same as Fig.~10 but for $l=3$ and $m=\pm 1$.}
\end{figure*}

\begin{figure*}
\includegraphics[width=1\textwidth]{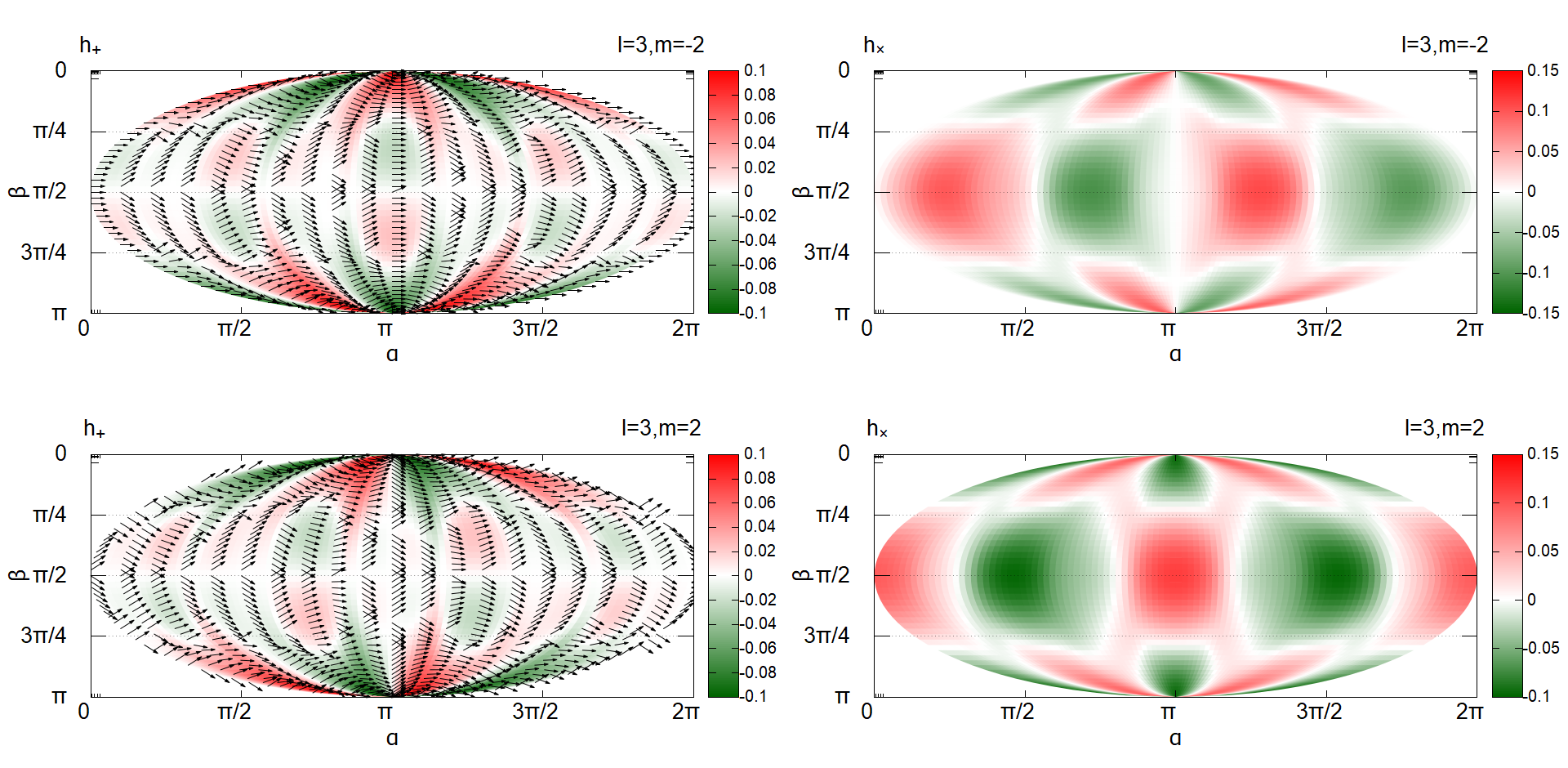}
\caption{\label{fig:wide}Same as Fig.~10 but for $l=3$ and $m=\pm 2$.}
\end{figure*}

\begin{figure*}
\includegraphics[width=1\textwidth]{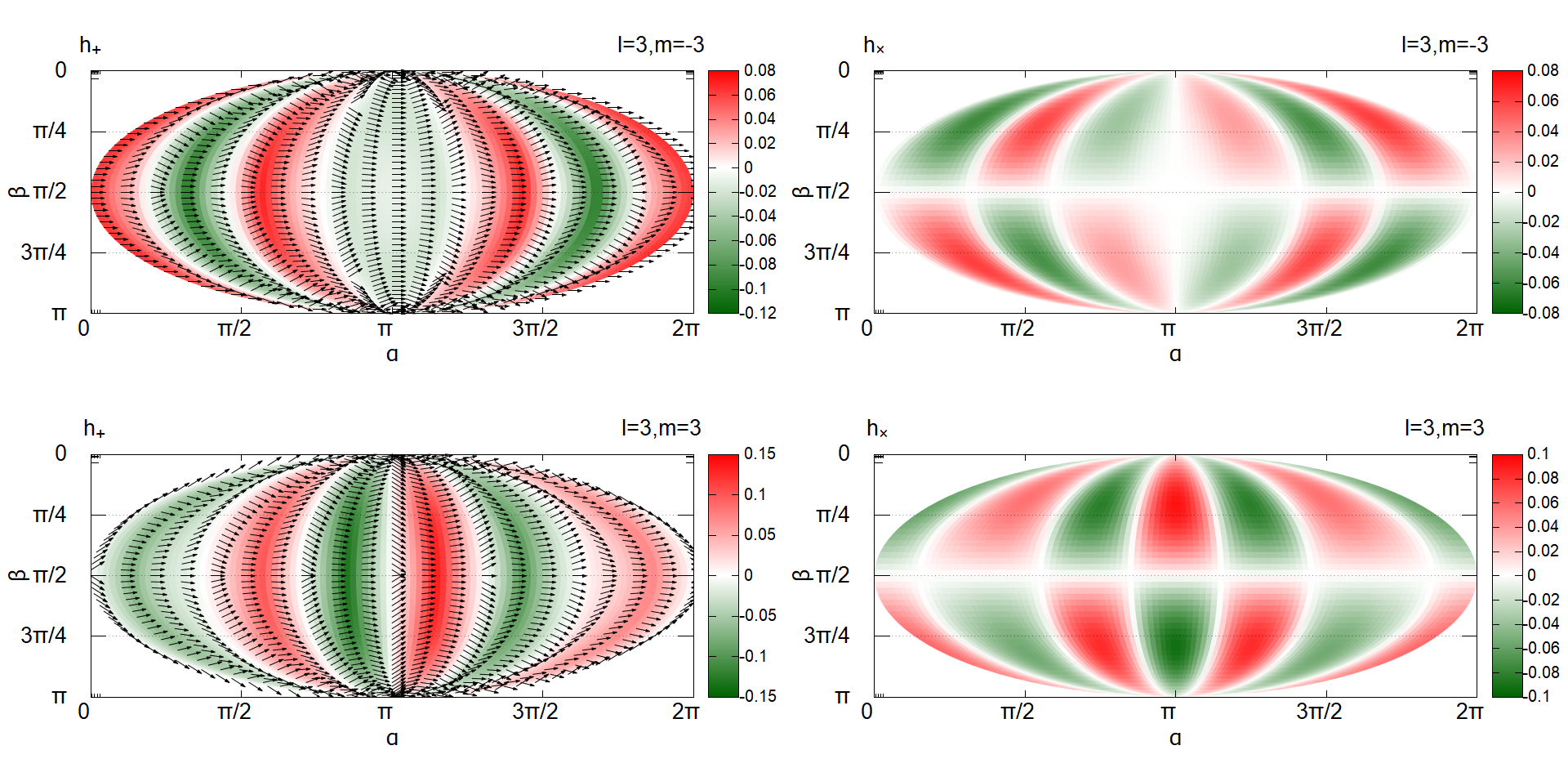}
\caption{\label{fig:wide}Same as Fig.~10 but for $l=3$ and $m=\pm 3$.}
\end{figure*}

\subsection{Dependence on the observer position: $\Psi_{lm}^{S}(\alpha, \beta)$ and polarizations } \label{KeepUp}
\noindent Now we shift our attention to $\Psi_{lm}^{S}$, the angular dependence of GW emissions, for models 1 and 2.\\
We begin with the axisymmetirc cases with $m=0$ in the spherical harmonics expansion. Since there is no dependence on the azimuth angle $\alpha$, we set it to zero. We show in Fig.~9 $\Psi_{lm}^{S}$ as a function of $\beta$. It turns out that $\Psi_{10}^{S}$ is vanishing identically. The dipolar asymmetry of neutrino emission hence does not produce GWs. As can be seen from the figure, the GWs cannot be detected if the observer is located either on the north pole or on the south pole. There are clearly preferential directions depending on $l$, e.g., the equator is the best direction for the $l = 2$ mode. Note also that the $\times$-mode of GW is vanishing in this axisymmetric case. As a result, the observed GW signals are linearly polarized irrespective of the detailed contents of harmonics in this case.\\
\noindent Next, we proceed to the non-axisymmetric case with $m\neq 0$. They are presented as color maps in Figs.~10 - 15 for all combinations of $l$ and $m$ up to the third order. Note that we also put arrows on top of the color maps for the +-mode that show the polarization patterns, which will be discussed later in this section.
This time not only the +-mode but also the $\times$-mode is nonvanishing; both modes are displayed as a function of $\alpha$ and $\beta$ in the Mollweide projection. As is evident, they have distinct features depending on the combination of $l$ and $m$. Of course, the actual signals are superpositions of these patterns.  Although it is not available for the moment, once a light curve of neutrino is provided as a function of solid angle by multi-dimensional computations of PNS cooling, we can immediately calculate such superpositions according to Eq.~(11). \\
\noindent In the non-axisymmetric case, the polarization of GW signals is non-trivial even for individual harmonics, since there exists the $\times$-mode in addition to the +-mode now. As mentioned just above, we are currently lacking the reliable estimate of the neutrino luminosity as a function of solid angle and we employ in this paper the simplest models, in which all harmonic components of the neutrino luminosity have the same time profile. Even in that case, the position angle of their linear polarization changes from position to position on the celestial sphere. As understood from Eq.~(11), it is determined by $\Psi_{lm}^{S}$ alone. As a matter of fact, under this circumstance, the GW waveform has a +-mode alone if one rotates the observer coordinates by an angle of $\Delta$ around the $Z'$-axis (see Fig.~1): 
\begin{equation}
\begin{split}
    \Delta &=\frac{1}{2}\arctan( \frac{h_{\times}(t)}{h_{+}(t)})\\
    &=\frac{1}{2}\arctan(\frac{\Psi_{\times}(\alpha, \beta)}{\Psi_{+}(\alpha, \beta)}).
\end{split}
\end{equation}
Here $\Delta$ is assumed to be in the range of $(-\frac{\pi}{2},\frac{\pi}{2})$.

\noindent We show in Figs.~10-15 this angle by the arrows attached to different points. Although shown in the panels for +-~modes alone for convenience, the polarization angle is determined by the ratio of the two modes (see Eq. (30)). The angles that the arrows make with the horizontal axis and are measured anticlockwise correspond to the values of  $\Delta$ above at their points. Note that the distributions of the value of $\Delta$ are anti-symmetric with respective to $\alpha=180^{\circ}$. It should be mentioned again that the actual signal is a superposition of all spherical harmonic modes and its polarization should be quite different from what we have shown here for the individual modes. We think, however, that this information is still useful. As a matter of fact, we do not know at the moment in what proportion these components are mixed in the total signal; in fact we simply "assumed" in this paper that the amplitudes of all components are 0.1 in models 1 and 2 and 0.01 in model 3, which is discussed in the next section. This is because we have no reliable simulations of PNS cooling in multi-dimensions at present as we repeatedly mentioned. In this sense, the analysis of each harmonic component is more important for the moment.

\subsection{Results of model 3} \label{KeepUp}
\begin{figure*}
\includegraphics[width=1.0\textwidth]{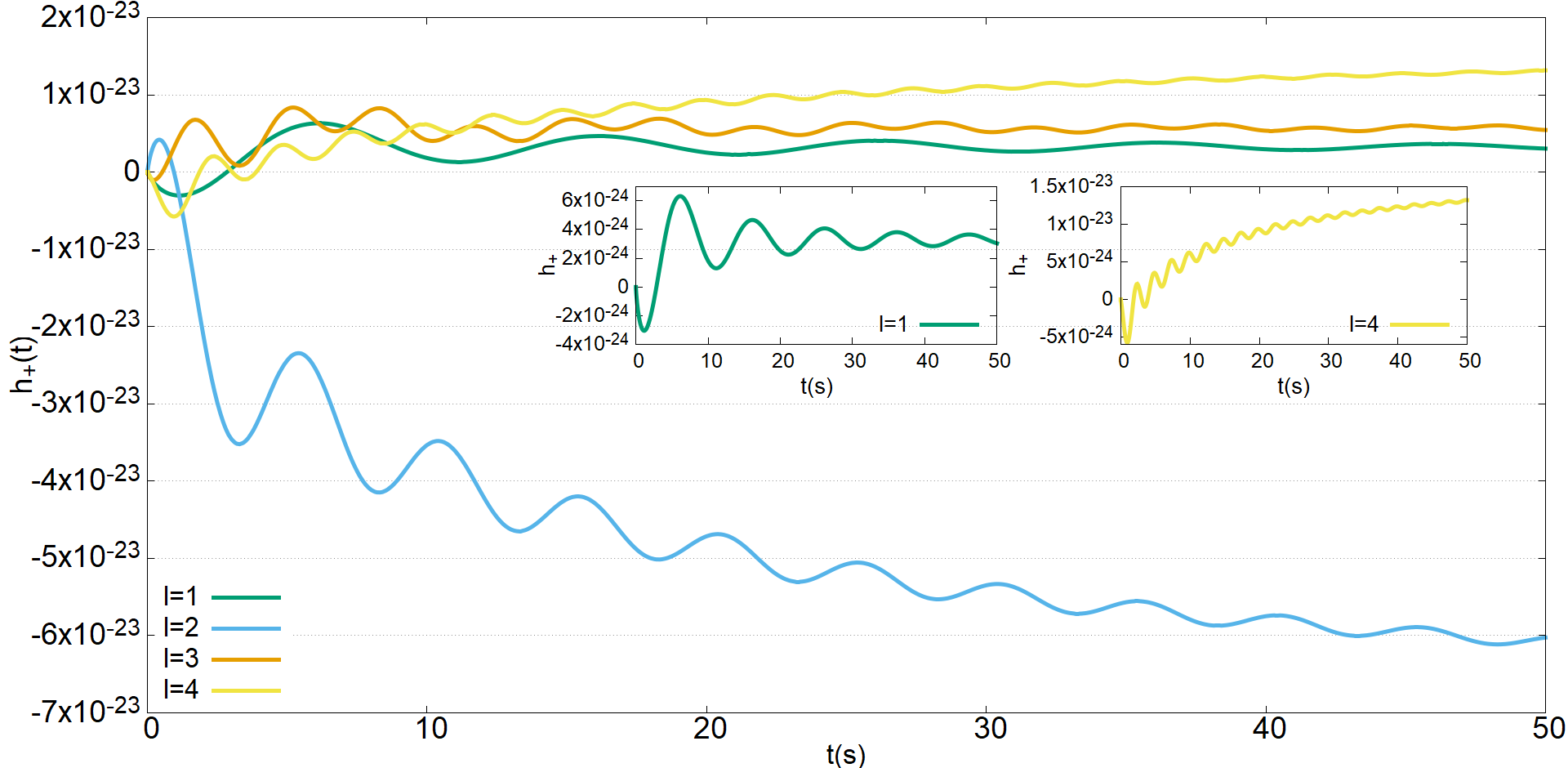}
\caption{\label{fig:wide}The gravitational waveform of the +-mode, $h_{+}(t)$, for different $l$'s. The position of observer is set to $\alpha=160^{\circ}, \beta=140^{\circ}$. The harmonic components are summed over $m$ for each $l$. We set $\epsilon=0.01$ and assume that the distance to the source is $R = 10$kpc. The insets are the zoom-ups of the harmonics $l =  1$ and $4$.}
\end{figure*}
\begin{figure*}
\includegraphics[width=1.0\textwidth]{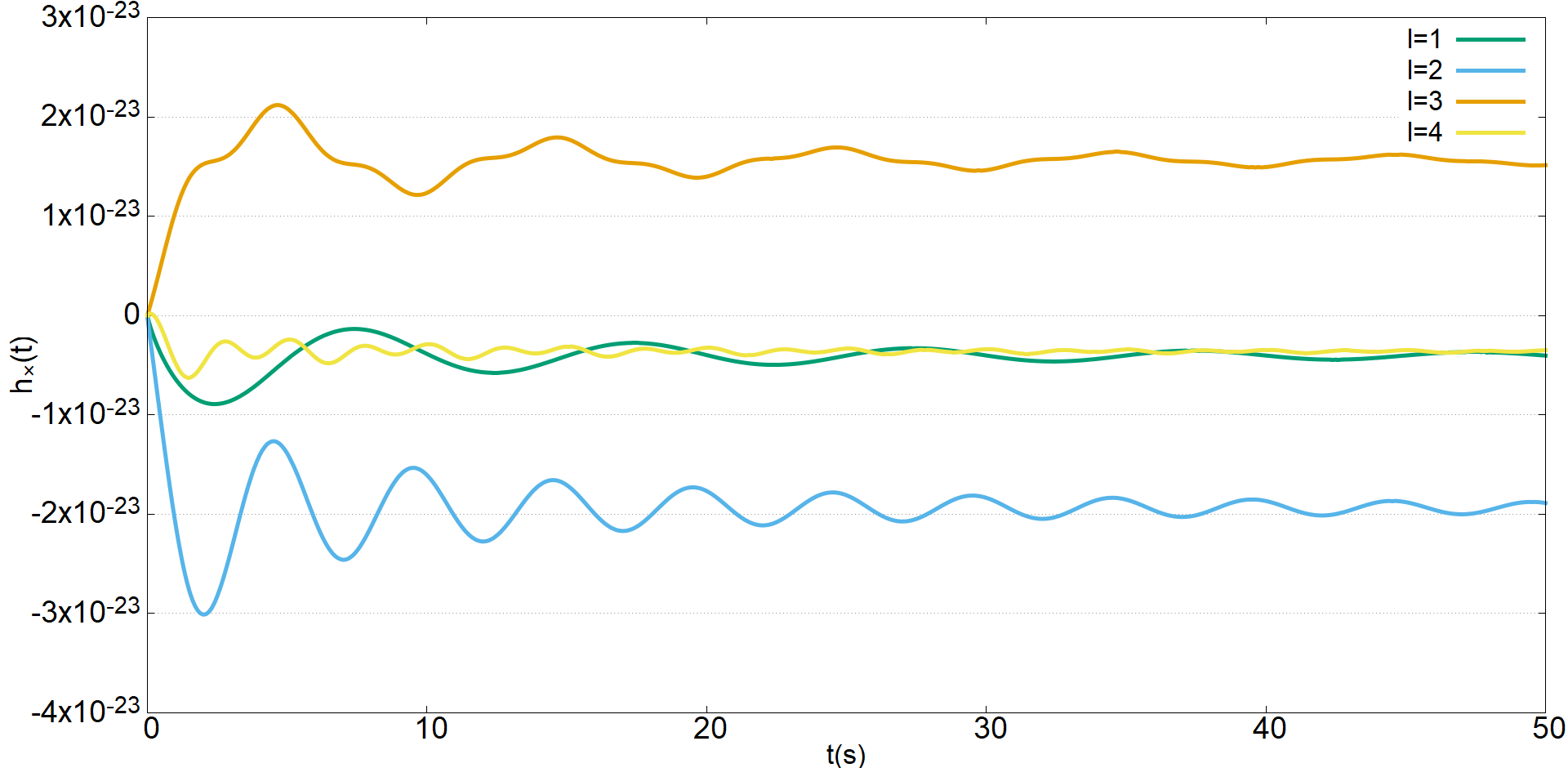}
\caption{\label{fig:wide}Same as Fig.~24 but for the $\times$-mode.}
\end{figure*}
\begin{figure}
\centering 
\includegraphics[width=0.5\textwidth]{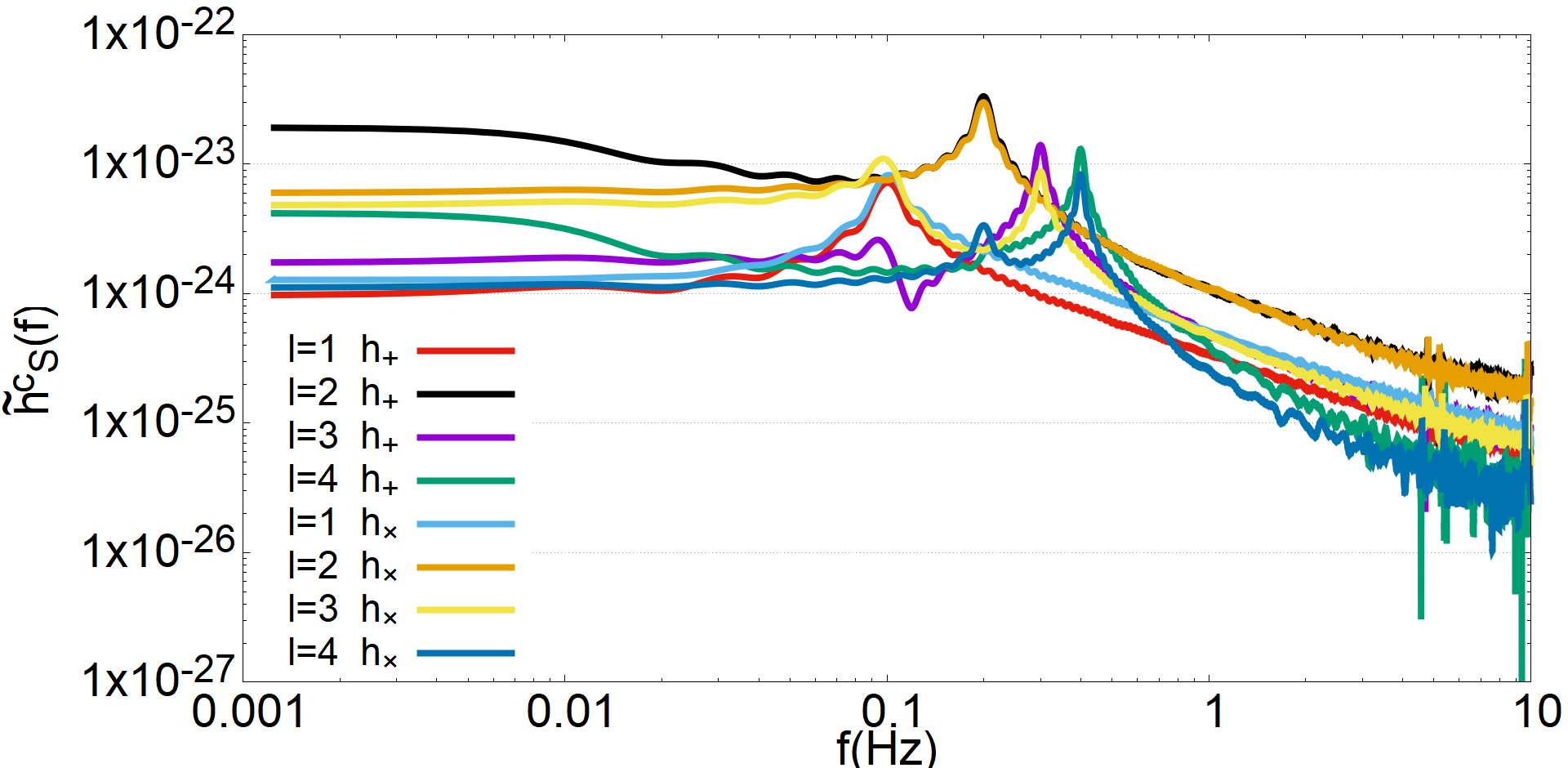}
\caption{The characteristic strains for model 3. Here $\Theta=\pi/2$. The harmonic components are summed
over $\it m$ for each $\it l$.}
\end{figure}
\begin{figure}
\centering 
\includegraphics[width=0.5\textwidth]{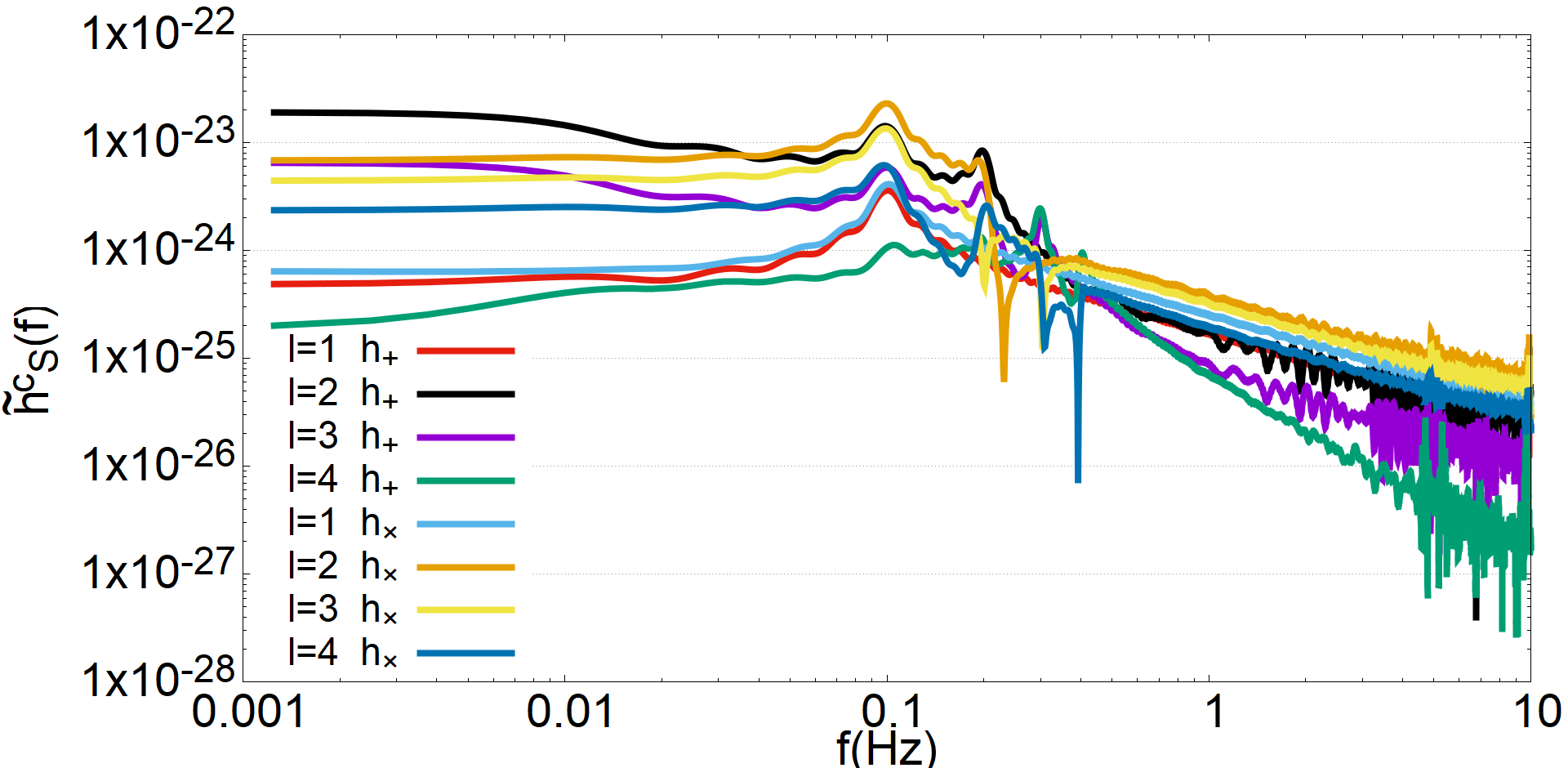}
\caption{Same as Fig.~26 but for $\Theta=\pi/6$.}
\end{figure}
\begin{figure*} 
\centering 
\includegraphics[width=1.0\textwidth]{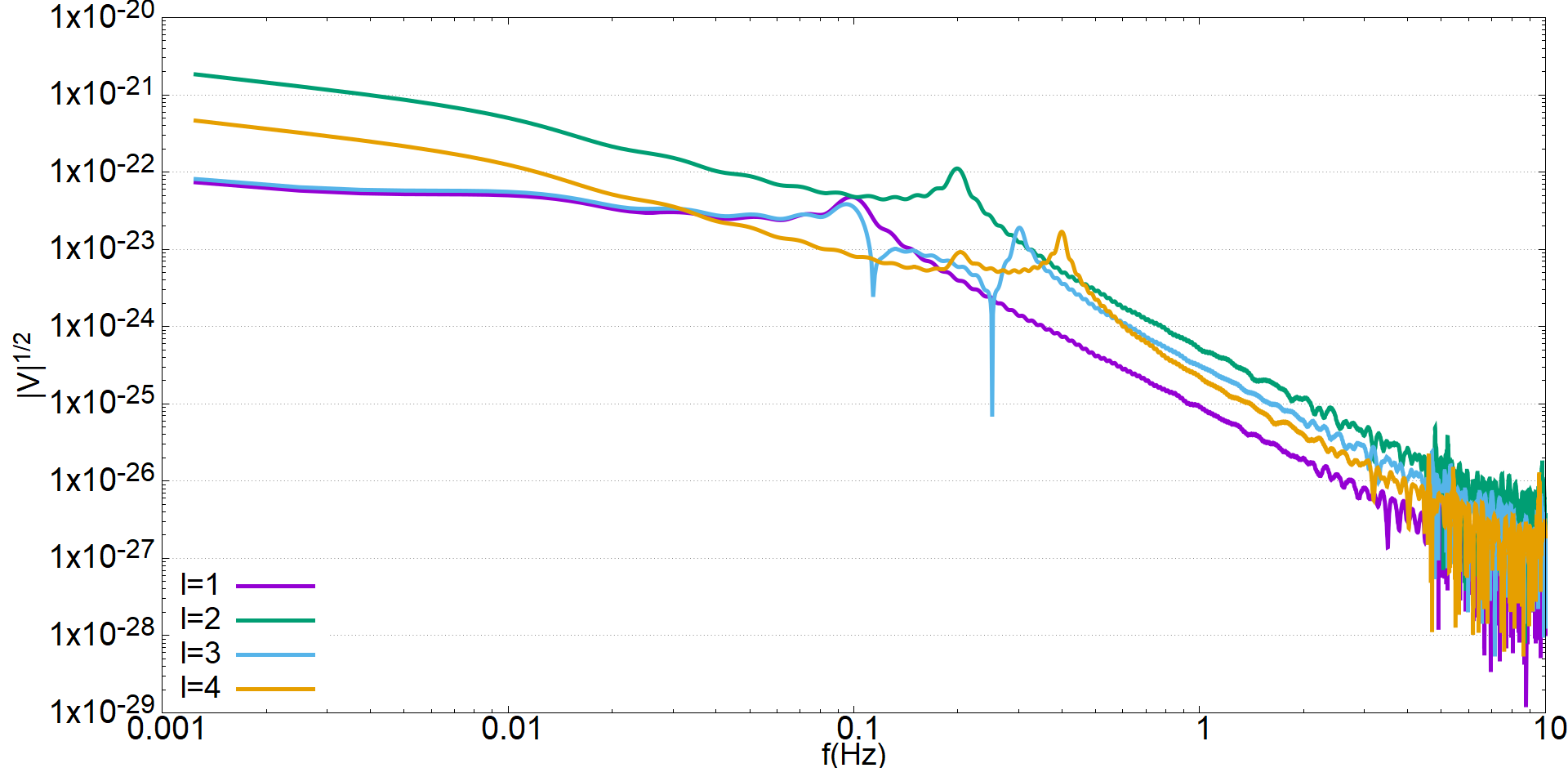}
\caption{\label{fig:wide}The square root of the absolute value of V as a function of $f$ for different $l$'s. The harmonic components are summed over $m$ for each $l$. The distance to the source is 10kpc.}
\end{figure*}
\begin{figure*} 
\centering 
\includegraphics[width=1.0\textwidth]{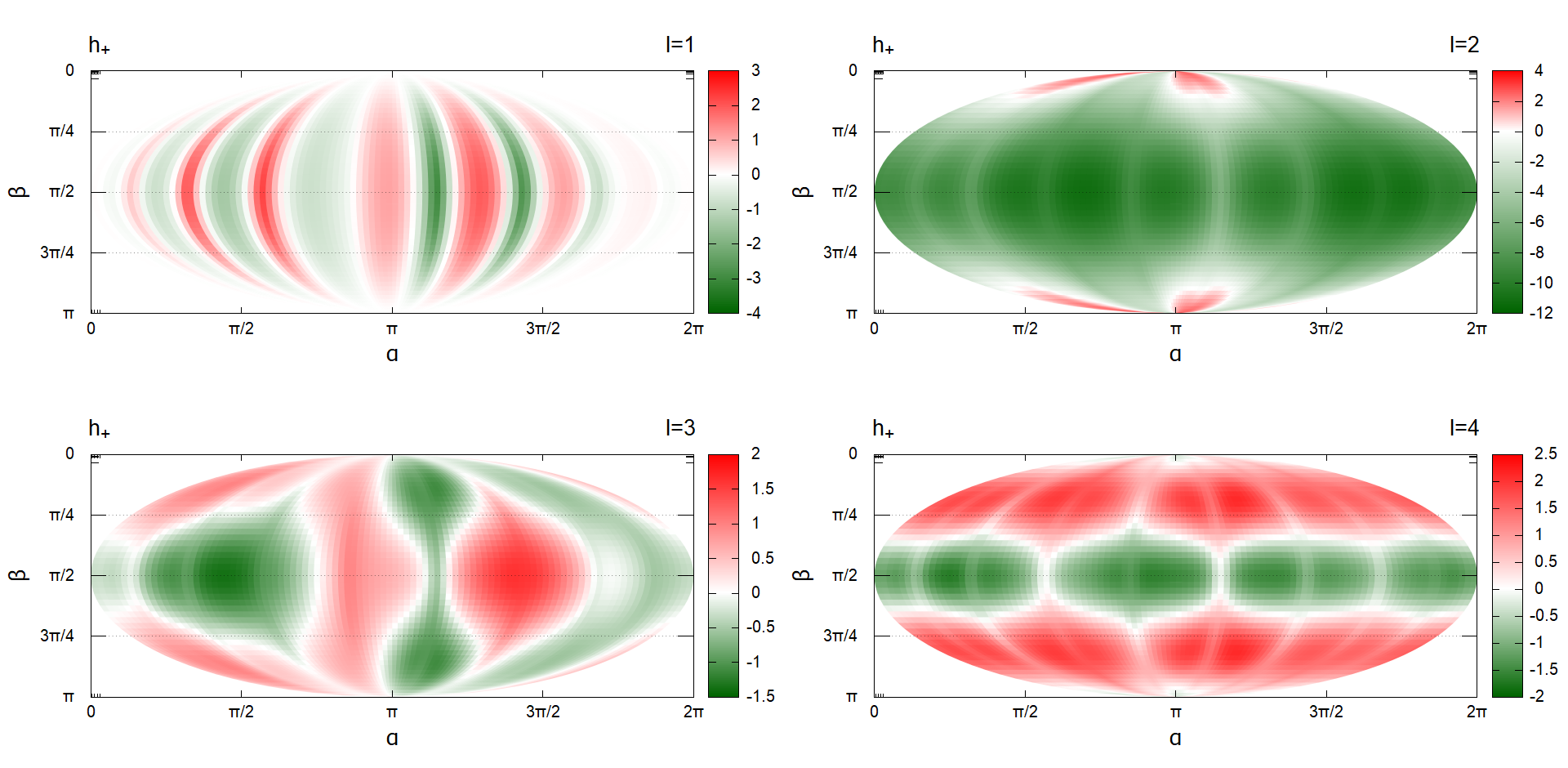}
\caption{\label{fig:wide}The amplitude of the +-mode  $h_{+}$ as a function of the observer position in the Mollweide projection at 20s for $l$ = 1 to 4. The harmonic components are summed
over $\it m$ for each $\it l$. The values in the color bar are in unit of  $10^{-22}$. The distance to the source is assumed to be 10kpc.}
\end{figure*}
\begin{figure*} 
\centering 
\includegraphics[width=1.0\textwidth]{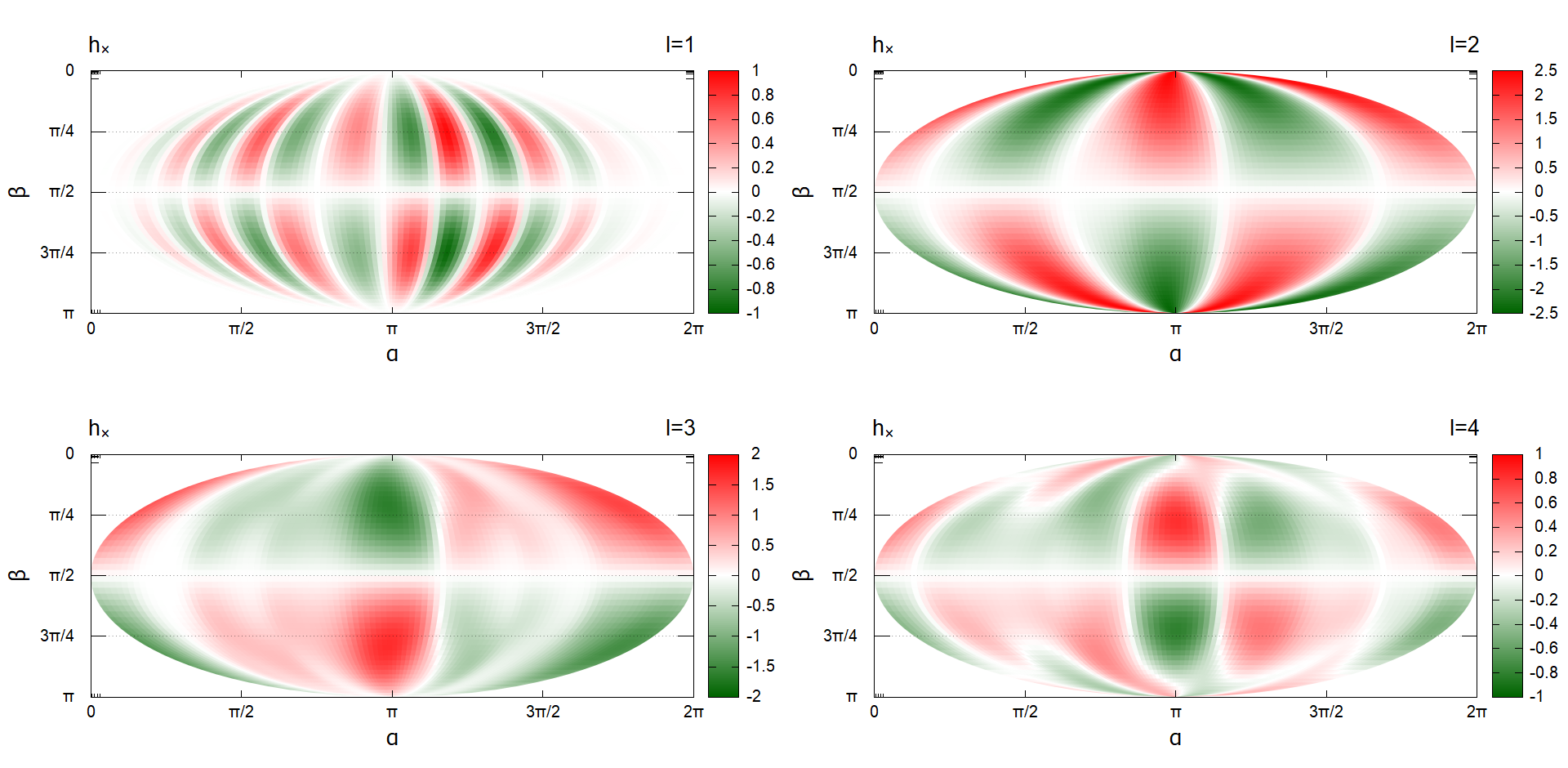}
\caption{\label{fig:wide}Same as Fig.~29 but for the $\times$-mode $h_{\times}$.}
\end{figure*}
\begin{figure*}
\centering 
\includegraphics[width=1\textwidth]{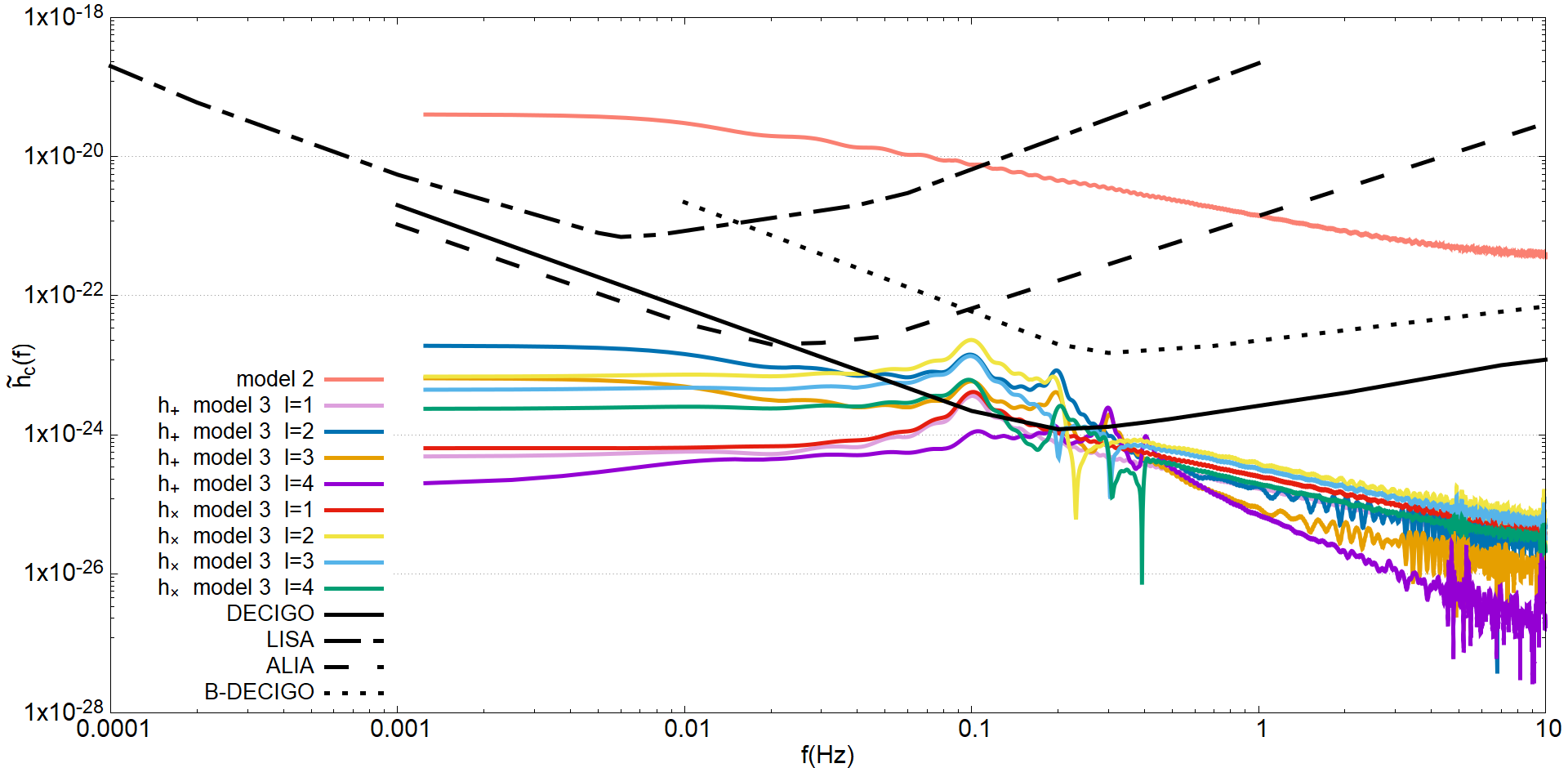}
\caption{The sensitivity curves for LISA, DECIGO, ALIA and B-DECIGO compared with the characteristic strains for models~2 and 3. We assume that the distance to the source is 10kpc and set $\epsilon=0.1$ and 0.01 for models 2 and 3, respectively.}
\end{figure*}

\noindent Now we proceed to model~3, in which the neutrino emission is axisymmetric with respect to a certain axis that itself is rotating at a constant angular frequency around another axis, which we identify with the $\it Z$-axis. As should be evident from  Eq.~(16), the spherical harmonics with $m \neq 0$ have their own time profiles that are different from that of the spherical component because of $\Phi(t)=\omega t=2\pi ft$ originating from the rotation of the symmetry axis around the $\it Z$-axis. This should be reflected in the GW signals.
In Figs.~16 and 17, we show the contributions to the waveforms $h_{+}(t,\alpha, \beta)$ and $h_{\times}(t,\alpha, \beta)$ from harmonics with different $l$'s up to 4. We assume that $\epsilon = 0.01$ and the observer is located at $\alpha=160^{\circ}$ and $\beta=140^{\circ}$ as a generic position and $\Theta = \pi / 2$. In these figures, we sum the harmonics  over $\it m$ for each $l$. This should be understandable, since the underlying angular distribution is axisymmetric with respect to a certain axis (see Eq.~(19)).\\
\noindent One finds that not only the +-mode but the $\times$-mode is also non-vanishing in this case. It is also clear in these figures that there are oscillatory features in the waveforms of individual  harmonic components. This is understandable from Eqs.~(8), (16), (19) and (20); $a_{lm}(t)$ contains
$\cos(m \omega t)$ or $\sin(m \omega t)$ in Eq.~(19), which originates from $\boldsymbol{Y}_{lm}(\Theta(t), \Phi(t))$ in Eq.~(16) (see also Eq.~(8)).
This is more clearly seen in the characteristic strain, which is presented for different harmonic components in Fig.~18. 
In fact, we expect peaks to appear in the characteristic strains at the frequencies of 
$mf$, in which $m$ satisfies $-l \leq
m \leq l$, since they are indices of the spherical harmonic function $Y_{lm}$. 
We found indeed a peak at $lf$ (i.e., $m = l$) in Fig. 18. We did not find, on the other hand, some of other peaks expected at $mf$ for $m \ne l$ in this figure: e.g. the peaks at $f$ and $3f$ (corresponding to $m = 1$, and $3$, respectively) for $l = 4$ are missing. This is because of the choice of $\Theta=\pi/2$.
As a matter of fact, we find from the spherical harmonic functions that
$h_{lm}^{amp}=0$ at odd (even) $m$'s for an even (odd) $l$ at this value of $\Theta$.
From this argument, we expect for $l = 4$ the peaks corresponding to $m = 2$ and 4 to show up and they do indeed although the former peak (corresponding to $m=2$) is not very clear. Note that the characteristic strain depends on $\Psi_{lm}$, which is in turn affected by the value
of $\Theta$. This is also the reason why some peaks are not remarkable as for the combination of $l = 4$ and $m = 2$. In order to vindicate
this interpretation, we look into the case with $\Theta = \pi/6$ in Fig. 19. We find indeed all the peaks at $mf$ with $0 < m \leq l$.
These features can be used to infer the rotation frequency $f$, which is supposed to be the frequency of the stellar rotation itself. Since the actual characteristic strain is a superposition of these individual spherical harmonic components, we will observe a few peaks at integral multiples of $f$. Then we may read out the rotation frequency as the greatest common divisor of the peak frequencies.\\
Next we move on to the polarization for this model. Since the neutrino beam is rotating, we expect that the GW signal will have a circular-polarization component. In order to see this quantitatively, we investigate the Stokes parameters\cite{Conneely_2019}
\cite{Hayama_2016}defined as 
\begin{align}
    &I(f,\widehat{k})=\langle|h_{+}(f,\widehat{k})|^{2}\rangle+\langle|h_{\times}(f,\widehat{k})|^{2}\rangle,\notag\\
    &Q(f,\widehat{k})=\langle|h_{+}(f,\widehat{k})|^{2}\rangle-\langle|h_{\times}(f,\widehat{k})|^{2}\rangle,\notag\\
    &U(f,\widehat{k})=2\langle \rm{Re}(h_{+}(f,\widehat{k})h_{\times}^{\ast}(f,\widehat{k}))\rangle,\notag\\
    &V(f,\widehat{k})=2\langle \rm{Im}(h_{+}(f,\widehat{k})h_{\times}^{\ast}(f,\widehat{k}))\rangle,
\end{align}
where I is the total intensity, Q and U stand for the linear polarization, and V represents the circular polarization, the focus here. In Fig.~20, we show the square root of the modulus of V as a function of frequency for different harmonic components.
We find again for each line that there are peaks at the frequencies, at which we observed the peaks in the characteristic strains above. If non-vanishing values of the V parameter with these peaks are observed, it may be an indication that the neutrino beam, and hence the PNS itself, are rotating at the frequency inferred from the peaks.\\
The anisotropy of GW emissions for this model is time-dependent as should be understood from Eq.~(21). We show, as an example, $h_+$ and $h_{\times}$  at t=20s as a function of the solid angle in the Mollweide projection in Figs.~21 and 22, respectively. Again we sum the harmonic components with the same $l$ over all $m$. Hence the results may be understood as the sum of anisotropies for different $m$'s shown in Figs.~10-15.

\subsection{Detectability} \label{KeepUp}
\noindent Finally, we compare in Fig.~23 the characteristic strains obtained so far with the planned sensitivities of prospective deci-Hertz GW detectors\cite{izumi2021detection}:
LISA\cite{baker2019laser}, DECIGO\cite{YAGI_2013}\cite{Hou_2021}\cite{Kawamura_2008}, ALIA\cite{Gong_2015}, B-DECIGO\cite{Isoyama_2018}; for model 2, the angular factor, $\Psi$, is not included whereas for model 3 it is included and the observer is assumed to be located at 
$\alpha = 160^{\circ}$ and $\beta = 140^{\circ}$;
the distance to the source is assumed to be 10kpc for both cases; the amplitudes of anisotropic neutrino emissions are $\epsilon$ = 0.1 and 0.01 for models 2 and 3, respectively. Note that it is not our goal in this paper to address the detectability of the GW signals with the detector noises being taken into account in detail. Our models are just not realistic enough for that. Instead we would like to give a proof of principle that there is a fair chance to detect the GW signals from the neutrinos emitted asymmetrically from PNS by some prospective satellite-borne GW detectors, which could give us an invaluable opportunity to get some information on the asymmetry and rotation of PNS. For that purpose, just a comparison of the characteristic strains with the sensitivity curves is enough. And note also that a mere detection does not imply immediately that we can extract relevant information well.
\\
One finds that the DECIGO has a potential to detect the gravitational wave signal for model 2 if the observer is located at a favorable position. In fact, the observed signal covers a wide range of frequency from $10^{-3}$ to 10 Hz, which will be sufficient to find the memory.
B-DECIGO, the precursor mission of DECIGO and hence its scaled-down version, has a lower sensitivity. It may still be able to detect the GW signal at its best-sensitivity frequency of $\sim$0.1Hz. The detection may be limited down to 0.01Hz.  
ALIA will be sensitive to a bit lower frequencies. It is better indeed than DECIGO at $\lesssim 0.01$Hz. LISA has its best sensitivity at even lower frequencies, $\sim 5\times 10^{-3}$Hz and may be able to detect the GW signal at these frequencies. Note, however, that the characteristic strain considered here does not include the angular factor $\Psi$ (Eq.~(13)) but only corresponds to the time evolution part Eq.~(12); the actual detectability depends on the observer's position. The signal to noise ratio\cite{Flanagan_1998} of this characteristic strain for DECIGO is crudely estimated as SNR $\approx 10^{3}$ in the frequency range of 0.001Hz $\sim$ 10Hz. Since the angular factor is typically $\Psi \approx 10^{-2} \sim 10^{-1}$ for the optimal direction (see Figs.~9, 10-15), the actual SNR may be $10 \sim 100$ for model~2.\\
In model 3, where the rotating (axisymmetric) neutrino beam is considered, the amplitude of anisotropy in the neutrino emission is assumed to be $\epsilon = 0.01$, one order of magnitude smaller than in model 2. As a result, the characteristic strains are smaller accordingly. Note also that they include the angular factor $\Psi$ in this case. The observer is assumed to be located at $\alpha  = 160^{\circ}$ and $\beta = 140^{\circ}$.
As is evident, the GW signals, the individual harmonics of which are shown in this figure, will be visible only to DECIGO. This is not only thanks to its best sensitivity but also due to the fact that its sensitivity is best at the frequencies, a few deci Hz, which coincide with the positions of the peaks in the GW signals. This coincidence obtains in turn because the rotation of PNS is assumed to be very slow as is the case for magnetars. Note again that the GW amplitude depends on the observer position; it is also affected by the angle $\Theta$ between the two axes, which is assumed to be $\pi/2$ in this figure. In fact the peak amplitude for $l = 2$ is smaller nearly by a factor of 2 to 3 for $\Theta = \pi / 6$ (see Figs.~18, 19). Nonetheless it is nice that DECIGO has a fair chance to detect such a tiny anisotropy in neutrino emissions and find the frequency of PNS rotation. Again just for reference we give here the crude estimation of the SNRs for DECIGO: $\sim 30$ for $\Theta = \pi /2$ and $\sim 20$ for $\Theta = \pi / 6$.\\

\section{\label{sec:level1}Summary}
\noindent We have investigated the gravitational radiation from the neutrinos emitted anisotropically from proto-neutron star in its cooling phase. By approximating with the piecewise exponential function the neutrino light curve obtained with the 1-D simulation of PNS cooling under spherical symmetry, we have identified 5 phases, which have their own decay time scales. 
We have employed this piecewise exponential function as well as the original light curve as the reference models for the time profiles of the anisotropic component of the neutrino emissions, since there is no multi-dimensional numerical simulation of PNS cooling available for the moment to provide quantitatively reliable data. We have also modified the PEF by hand to study the effect of possible convection in PNS.
We have considered yet another model, in which neutrinos are emitted axisymmetrically with respect to the axis that is rotating around a space-fixed axis. We have in mind a strongly magnetized PNS, or a magnetar, with a magnetic dipole moment misaligned with the rotation axis. We assumed a slow rotation with a period of 10s as typified by observed magnetars\cite{Kaspi2010}.\\
We have then calculated the GW waveforms with our formula based on the spherical-harmonic expansion of the neutrino luminosity. The formula should be convenient once multi-dimensional simulations become available to provide the angle-dependent neutrino luminosities as a function of time in the near future. We have also calculated the characteristic strains for different spherical harmonic components, both axisymmetric ($m=0$) and non-axisymmetric ($m \ne 0$). We have demonstrated that some of the time scales that characterize the decay of luminosity in different phases may be derived from the bumps in the characteristic strain at low frequencies for the cooling without convection; for more rapid cooling in the presence of convection, we may obtain some of the cooling times either from the locations of slope changes or from the oscillation periods in the characteristic strain.\\
\noindent For the axisymmetric neutrino emissions, the $\times$-mode always vanishes and the GW signal is linearly polarized just as for those from axisymmetric matter motions. 
For non-axisymmetrically emitted neutrinos, on the other hand, the $\times$-mode is nonvanishing anymore. We have calculated for each harmonic mode the polarization angle as a function of the observer position on the celestial sphere. It is actually correlated with the pattern of GW emissions, since both of them are dictated by the angular factor $\Psi$ in Eq.~(13).
In the absence of reliable estimates of the neutrino anisotropy in the proto-neutron star cooling at present, all we can do is to investigate the behaviours and properties of each spherical harmonic mode individually. Note that once we obtain a reliable theoretical prediction of the anisotropic neutrino emissions from the proto-neutron star, it is easy for us to obtain the observed gravitational wave signals just by superimposing these results on top of each other with appropriate weights. These results are hence expected to guide the near-future detection of deci-hertz gravitational waves.\\
Then we have applied the same analysis to the model meant for the rotating magnetar with its magnetic axis misaligned with the rotation axis. The rotation of the neutrino beam is indeed reflected as the peaks in the characteristic strain and as the nonvanishing Stokes V parameter. We have observed that these peaks appear at the frequencies of integral multiples of the rotation frequency, the fact that we may be able to employ to extract from the GW signals the information on the PNS rotation at its birth. \\
Finally, we have compared the characteristic strains obtained in these models with the sensitivity curves of some prospective deci-hertz detectors such as LISA, DECIGO, ALIA and B-DECIGO. We have shown that DECIGO has a potential to detect all the features mentioned above, covering a frequency range best suited for the GW signals considered in this paper and having a high enough sensitivity.\\
Our models are admittedly very crude. We assumed the same time profile for all harmonic components, which is certainly not true. The amplitudes of anisotropy in neutrino emissions adopted in this paper may not be typical. We have to wait for a realistic simulation of PNS cooling in multi-dimensions, though. Once it is done, however, we are ready to calculate the low-frequency GW signal from  the results they provide.

\section{Acknowledgment}
\noindent L.F. gratefully acknowledge useful discussions with K. Sugiura, H. Suzuki, T. Morinaga and W. Iwakami. He is also owing K. Sugiura for data of his 1-D simulation of of PNS cooling. This work was supported in part by Grants-in-Aid for Scientific Research (21H01083) and the Grant-in-Aid for Scientific Research on Innovative areas
"Unraveling the History of the Universe and Matter Evolution with Underground Physics" (19H05811) from the Ministry of Education, Culture, Sports, Science and Technology (MEXT), Japan. S. Y. is supported by the Institute for Advanced Theoretical and Experimental Physics, and Waseda University and the Waseda University Grant for Special Research Projects (project number: 2020C-273, 2021C-197).

\bibliography{apssamp}

\end{document}